\documentclass{aa}  
\usepackage{graphicx}
\usepackage{txfonts}
\begin{document}
   \title{The {\it XMM-Newton} Bright Serendipitous Survey\thanks{Based on 
observations collected at the Telescopio Nazionale Galileo (TNG) and
at the European
Southern Observatory (ESO) and on observations
obtained with {\it XMM-Newton}, an ESA science mission
with instruments and contributions directly funded by ESA Member States and 
the USA (NASA)}}

   \subtitle{Identification and optical spectral properties}

   \author{A. Caccianiga\inst{1}
          \and
	  P. Severgnini\inst{1}
          \and
          R. Della Ceca\inst{1}
          \and
          T. Maccacaro\inst{1}
          \and
	  F. Cocchia\inst{1,2}
	  \and
	  X. Barcons\inst{3}
	  \and
          F. J. Carrera\inst{3}
	  \and
	  I. Matute\inst{6}
	  \and
	  R. G. McMahon\inst{4}
	  \and
          M. J. Page\inst{5}
	  \and
	  W. Pietsch\inst{6}
	  \and
	  B. Sbarufatti\inst{7}
	  \and
	  A. Schwope\inst{8}
	  \and
	  J. A. Tedds\inst{9}
	  \and
	  M. G. Watson\inst{9}
	  %\fnmsep\thanks{} 
          }

   \offprints{A. Caccianiga}
   \institute{INAF - Osservatorio Astronomico di Brera, via Brera 28, 
 I-20121 Milan, Italy\\
              \email{alessandro.caccianiga@brera.inaf.it}
         \and
INAF - Osservatorio Astronomico di Roma, via di Frascati 33, 00040 Monte 
Porzio Catone, Italy 
\and
Instituto de F\'\i sica de Cantabria (CSIC-UC), Avenida de los
Castros, 39005 Santander, Spain
\and
Institute of Astronomy, Madingley Road, Cambridge CB3 0HA, UK
\and
           Mullard Space Science Laboratory, University College London, 
      Holmbury St. Mary, Dorking, Surrey, RH5 6NT, UK
   \and
Max-Planck-Institut f\"ur extraterrestrische Physik, Giessenbachstrasse, 
85741 Garching, Germany 
\and
INAF - IASFPA, via Ugo La Malfa 153, 90146 Palermo, Italy
\and
      Astrophysikalisches Institut Potsdam (AIP), 
      An der Sternwarte 16, 14482 Potsdam, Germany
      \and
     X-ray \& Observational Astronomy Group, Department of Physics and Astronomy, 
     Leicester University, Leicester LE1 7RH, UK
         %\an
          %  ...\\
           %  \email{}
            % \thanks{}
            }

   \date{}

% \abstract{}{}{}{}{} 
% 5 {} token are mandatory
 
  \abstract
   {}
   {We present the optical classification and redshift of 348 X-ray 
selected sources from the XMM-Newton Bright Serendipitous Survey (XBS)
which contains a total of 400 objects (identification level = 87\%). About
240 are new identifications. In particular, we  discuss in detail the  
classification 
criteria adopted for the Active Galactic Nuclei population.}
   {By means of systematic spectroscopic campaigns and through the literature 
search we have collected an optical spectrum for the large majority of the
sources in the XBS survey and applied a well-defined 
classification ``flow-chart''. }
   {We find that the AGN represent the most numerous population  at the 
flux limit of the XBS survey ($\sim$10$^{-13}$ erg cm$^{-2}$ s$^{-1}$) 
constituting 80\% of the  XBS sources selected in the 0.5-4.5 keV energy 
band and 95\% of the ``hard'' (4.5-7.5 keV) selected objects. Galactic 
sources populate significantly the 0.5-4.5 keV sample (17\%) and only 
marginally (3\%) the 4.5-7.5 keV sample. The remaining sources in both samples 
are clusters/groups of galaxies and normal galaxies (i.e. probably 
not powered by an AGN). Furthermore, the percentage of type~2 AGN (i.e. 
optically absorbed AGNs with A$_V>2$mag) dramatically increases 
going from the 0.5-4.5 keV sample 
(f=N$_{AGN 2}$/N$_{AGN}$=7\%) to the 4.5-7.5 keV sample (f=32\%). 
We finally propose two simple diagnostic plots that can be easily used to 
obtain the spectral classification  for relatively low redshift AGNs even 
if the quality of the spectrum is not good. 
%We finally find a good match between the optical classification and the
%X-ray absorption level inferred from the analysis of the hardness-ratios,
%with a dividing line between absorbed and unabsorbed AGN consistent with that
%expected from Galactic A$_V$/N$_H$ relation 
%(N$_H\sim$4$\times$10$^{21}$ cm$^{-2}$).
}
   {}

   \maketitle

%__________________________________________________________________

\section{Introduction}
In the last few years {\it XMM-Newton} and {\it Chandra} telescopes 
have represented an excellent tool to survey the hard X-ray sky  
at all fluxes, from relatively bright (10$^{-13}$ erg cm$^{-2}$ s$^{-1}$, e.g.
Della Ceca et al. 2004 and references therein), to medium 
(10$^{-13}$ erg cm$^{-2}$ s$^{-1}$-10$^{-14}$ erg cm$^{-2}$ s$^{-1}$, 
e.g. Barcons et al. 2007 and references therein)
and deep (10$^{-14}$-10$^{-16}$  erg cm$^{-2}$ s$^{-1}$, 
Brandt \& Hasinger 2005; 
Worsley et al. 2005 and references therein) fluxes. 
At the energies ($\sim$0.5-10 keV) covered  
by the instruments on board  these two telescopes, Active Galactic Nuclei 
(AGN) can be efficiently selected and studied even when affected by 
large levels of absorption (up to N$_H\sim$10$^{24}$ cm$^{-2}$, corresponding
to an optical absorption of A$_V\sim$500 mag). This important characteristic,
combined with the good/excellent spatial and energy resolutions of the
detectors, makes the ongoing surveys a fundamental tool for AGN studies.
At the same time, these new surveys represent an observational challenge
at wavelengths different from the X-ray ones: multiwavelength follow-ups 
of X-ray sources, particularly in the optical domain, are decisive to 
derive the distance and to understand the properties of the selected objects 
but they also require large fractions
of dedicated observing time at different telescopes. Probably one of the most
challenging and time-consuming efforts is the optical spectroscopic follow-up
of the selected sources.

One of the  primary goals of all these hard X-ray surveys is to explore 
the population of absorbed AGN and, to this end, an optical 
classification that can reliably separate between optically absorbed 
and non-absorbed objects is always required. 
Two important limits, however, affect the spectroscopic follow-ups of 
deep and, in part, medium surveys: first, 
the optical counterparts are often too faint to be spectroscopically
observed even at the largest optical telescopes currently available;
second, even when a spectrum can be obtained, its quality is not always
good enough to provide the critical pieces of information that are
required to assess 
a reliable optical classification. These two problems often  limit
the final scientific results that are based on the optical classification of
medium/deep surveys. 

On the contrary, bright surveys offer
the important possibility of obtaining a reliable optical classification
for virtually all (with some exceptions, as discussed in the next sections) 
the selected sources. 
The disadvantage of dealing with shallow/wide-angle samples is
that the techniques 
to observe efficiently many sources at once, like Multi-objects or 
fibers-based methods,  cannot be applied for the
optical follow-up, given the low space-density of sources at bright
X-ray fluxes. 
The only suitable method, the ``standard'' long-slit technique, 
requires many independent observing nights to achieve the completion of the 
optical follow-up. 

In this paper we present and discuss in detail the optical classification 
process of the {\it XMM-Newton} Bright Serendipitous Survey (XBS, 
Della Ceca et al. 2004), which currently represents the widest (in terms
of sky coverage) among the existing {\it XMM-Newton} or {\it Chandra} 
surveys for which  a spectroscopic follow-up has been almost completed.
The aim of the paper, in particular, is to
provide not only a generic classification of the sources and their redshift 
but also
a quantification, in the limits of the available data, of the corresponding
threshold in terms of level of optical absorption. 

The paper is organized as follows: in Section~2 we describe the XBS survey,
in Section~3 we describe the process of identification of the optical 
counterpart, in Section~4 and Section~5 we respectively  summarize our own 
spectroscopic 
campaigns carried out to collect the data as well as the data obtained from
the literature. In Section~6 we shortly discuss the data reduction and
analysis of the optical spectra and in Section~7 we give the details on
the classification criteria adopted for the sources in the XBS survey. 
In Section~8 we propose two diagnostic plots that can be used
to easily classify the sources into type~1 and type~2 AGN. 
The resulting catalogue is presented in Section~9 while in Section~10 
we briefly discuss the optical breakdown and the redshift distribution of
the sources. 
The conclusions are finally
summarized in Section~11 .  Throughout this paper H$_0$=65 km s$^{-1}$ 
Mpc$^{-1}$, $\Omega_{\Lambda}$=0.7 and $\Omega_M$ = 0.3 are assumed.

 \section{The XMM-{\it Newton} Bright Serendipitous Survey}
The XMM-{\it Newton} Bright Serendipitous Survey (XBS survey, 
Della Ceca et al. 2004)
is a wide-angle ($\sim$28 sq. deg) high Galactic latitude ($|b|>$20 $\deg$) 
survey based on the XMM-{\it Newton} 
archival data. It is composed of two samples both flux-limited 
($\sim$7$\times$10$^{-14}$ erg cm$^{-2}$ s$^{-1}$) in two separate energy 
bands: the ``soft'' 0.5-4.5 keV band (the BSS sample) and the hard 4.5-7.5 keV 
band (the HBSS sample). A total of 237 (211 for the HBSS sample) 
independent fields have been used to select   400 sources, 
389 belonging to the BSS sample and 67 to the HBSS sample (56 sources are
in common). The details on the fields selection strategy, the sources
selection criteria and the general properties of the 400 objects 
are discussed in Della Ceca et al. (2004). 

One of the main goals of the survey is to provide a well-defined and 
statistically complete census of the AGN population with particular 
attention to the problem of obscuration. To this end, the possibility 
of  comparing X-ray and optical spectra of good quality 
for all the sources present in the two complete samples offers 
a unique and fundamental tool to statistically study  the effect of absorption
in the AGN population in an unbiased way.
Indeed, most of the X-ray sources of the XBS survey 
have been detected with enough counts to allow a reliable X-ray  
spectral analysis. At the same time, most of the sources 
have a relatively bright (R$<$22 mag, see next section) 
optical counterpart and they can be 
spectroscopically characterized using a 4-meters-class telescope.

To date,  the spectroscopic identification level has reached 
87\% (87\% and 97\% considering the BSS and the HBSS samples separately). 
The results of the spectroscopic campaigns are discussed in the following
sections.

\section{Identification of the optical counterpart}

   \begin{figure}
   \centering
    \includegraphics[width=9cm]{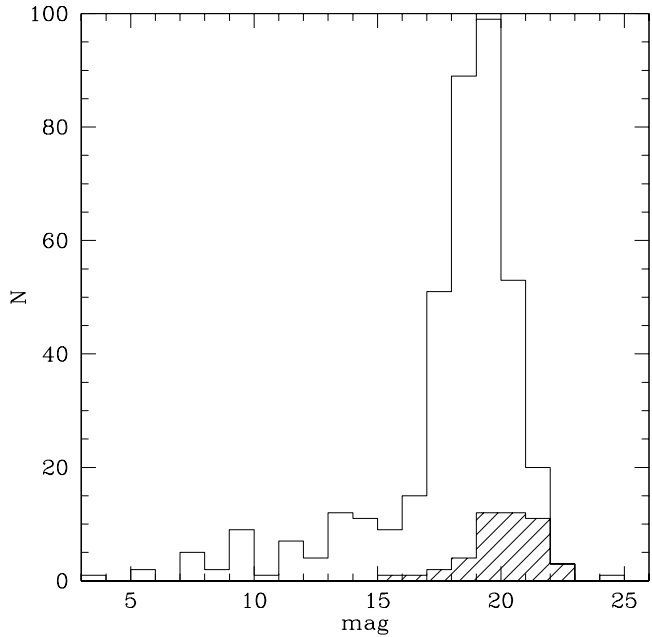}

   \caption{Magnitude distribution of the XBS optical counterparts. Most (94\%)
of the magnitudes are in a red filter. Shaded histogram represents the sources
without a spectral classification.
}
              \label{mag_dist}
    \end{figure}

The identification of the optical counterparts of the XBS sources is 
relatively easy given the combination of the good positions of the 
XMM-{\it Newton} sources (90\%\ error $\sim$4\arcsec, Della Ceca et al. 2004) 
and the brightness of the sources: X-ray sources with F$_X>$10$^{-13}$ 
erg cm$^{-2}$ s$^{-1}$ are expected to have an optical counterpart
brighter than 22 mag for X-ray-to-optical flux ratios below 20 (i.e. for
the majority of type~1 AGN, galaxies and stars). Only the rare 
(but interesting)
sources with extreme X-ray-to-optical flux ratios, like the distant type~2 QSOs
(e.g. Severgnini et al. 2006), are expected to have magnitudes as faint as 
R$\sim$25. For this reason, for the large majority of the XBS sources we
have been able to unambiguously pinpoint the optical counterpart using the
existing optical surveys (i.e. the 
DSS I/II\footnote{http://stdatu.stsci.edu/dss/} and the 
SDSS\footnote{http://www.sdss.org/}). 
In particular, we have found the optical counterpart of about 88\% of 
the XBS sources on the DSS with  
a red magnitude (the APM\footnote{http://www.ast.cam.ac.uk/$\sim$apmcat/} 
red magnitude) brighter than $\sim$20.5. 
All but 6 of the remaining sources 
have been optically identified either through dedicated photometry 
or using the SDSS catalogue. The red magnitudes of these sources 
are relatively bright (R between 20.5 and 22.5) with one exception: an 
R=24.5 object (XBSJ021642.3-043553), that turned out to be a distant (z=1.985)
type~2 QSO (Severgnini et al. 2006). 
For 6 objects we have not yet found the optical counterpart but only
for two of these we have relatively  deep images that have 
produced a faint lower limit on the R magnitude (R$>$22.8  and R$>$22.2
respectively). For the other 4 sources we only have the upper limit
based on the DSS plates.
   \begin{figure}
   \centering
    \includegraphics[width=9cm]{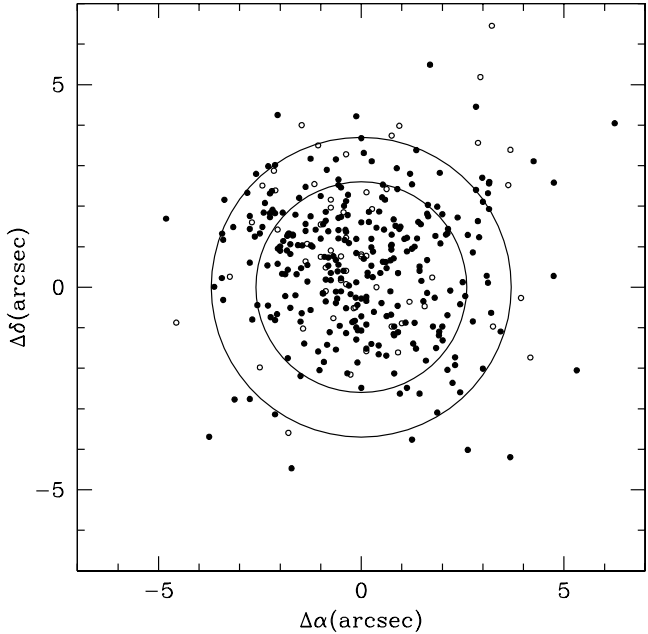}

   \caption{X-ray/optical positional offsets of the 348 
XBS sources with a spectral classification. Open circles are
stars and clusters of galaxies while filled points are the 
remaining sources (AGNs and ``elusive AGN candidates''). The two circles
represent the regions including 68\% and 90\% of the points}
              \label{opt_posit}
    \end{figure}

   \begin{figure}
   \centering
    \includegraphics[width=9cm]{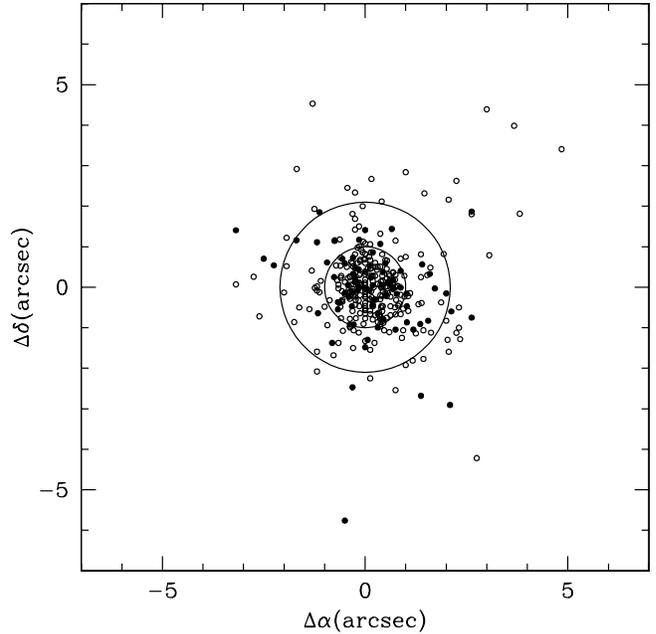}

   \caption{X-ray/optical positional offsets of the
XBS sources  in common with the
2XMM catalogue (343 in total). In this case the X-ray positions are taken
from the 2XMM catalogue that has been produced with recent versions
of the Standard  Analysis System. Symbols and regions 
as in Fig.~\ref{opt_posit}}
              \label{newopt_posit}
    \end{figure}

In conclusion, we have found the most-likely optical counterpart in the large
majority of the 400 XBS sources (all but 6 sources). The magnitude distribution
of the counterparts is presented in Fig.~\ref{mag_dist}. Since we have not
carried out a systematic photometric follow-up of the XBS objects, we do 
not have an homogeneous set of magnitudes in a well defined filter. 
In Fig.~\ref{mag_dist} we have reported the magnitudes either from existing 
catalogues (e.g. APM, SDSS, NED\footnote{http://nedwww.ipac.caltech.edu/}, 
Simbad\footnote{http://simbad.u-strasbg.fr/simbad/}) or from our own 
observations. Most of them (94\%) are in a red filter while the remaining 
6\% (all bright stars with mag$<$13) are in V or B filters.

In Fig.~\ref{opt_posit} we show the X-ray/optical 
positional offsets of the 348 XBS sources discussed in this paper 
(i.e. those with a spectral classification). All the identifications
have offsets below $\sim$7$\arcsec$, with the majority  ($\sim$90\%) of 
sources having an offset below 3.8$\arcsec$. In Fig~\ref{opt_posit} we have 
distinguished the objects spectroscopically classified as stars and clusters
of galaxies (indicated with open circles) from the rest of the sources
(filled circles) since both stars and clusters may suffer from larger
positional offsets due to the presence of proper motions (stars) or, in the 
case of clusters of galaxies, due to the 
intrinsic offset between the X-ray source (the intracluster gas) and the
optical object (e.g. the {\it cD} galaxy). Indeed, the circle including 90\% 
of stars and clusters is larger ($\sim$4.5$\arcsec$) than the
circle computed using all the sources. 
%                                     One column figure (place early!)

In the last years
the XMM-{\it Newton} images have been reprocessed with improved versions 
of the SAS and the astrometry has been refined and corrected. We have 
thus recomputed the X-ray-optical offsets using the improved X-ray
positions included in the preliminary version of the 
second XMM-{\it Newton} Serendipitous EPIC Source Catalogue (2XMM, 
Watson et al. 2007, in preparation, see also http://xmm.vilspa.esa.es/xsa/). 
In Fig.~\ref{newopt_posit} we plot these newly computed offsets for the 
objects that are in common with the 2XMM catalogue. The improvement is
evident, with 90\% of the sources (excluding stars and clusters) having 
an offset below 2.1$\arcsec$. The sources with relatively large offsets
(4-5$\arcsec$) are mostly stars and clusters. 
All but 2 extragalactic ``non-clusters'' objects 
have X-ray-to-optical offsets below 4\arcsec.
By inspecting the X-ray images of the two extragalactic ``non-clusters'' 
objects with 
large offsets (XBSJ095054.5+393924, a type~1 QSO at z=1.299 
and XBSJ225020.2-64290, a type~1 QSO at z=1.25) we have found strong 
indications that both objects are
the result of a source blending which has ``moved'' the centroid of the
X-ray position between two nearby objects. Interestingly, in one of
these cases (XBSJ225020.2-642900) we have spectroscopically observed also the  
second (and fainter) nearby object and found a very similar spectrum
of type~1 QSO at the same redshift (1.25). This could either be a
real QSO pair or, alternatively, the result of gravitational lensing 
caused by a (not visible) galaxy.

In conclusion, excluding these two objects, for which the X-ray position is
not accurate, all the XBS sources classified as extragalactic objects 
have an optical counterpart within 4\arcsec\ using the improved X-ray positions
and 90\% have offsets within 2.1\arcsec. 

\subsection{Estimate of the number of spurious X-ray/optical associations}
As discussed above, the optical counterparts found for the XBS sources
have R magnitudes brighter than 22.5 (except for one object) 
with a large fraction (88\%) 
of them having magnitudes brighter than 20.5 (i.e. they are visible on the
DSS plates). Given the density of AGN at the magnitude limit of R=22.5 
(e.g. Wolf et al. 2003), the probabilty of finding 
an AGN by chance within 4$\arcsec$ from an unrelated X-ray source 
is $\sim$5$\times$10$^{-4}$  which translates
into an expected number of  $\sim$0.2  spurious AGN identifications in the 
entire XBS survey. Therefore it is
reasonable to consider all the objects optically classified as emission
line AGN (or BL Lac objects) as the correct counterparts of the X-ray sources.
Stars and galaxies, instead, may contaminate the identification process, 
given their higher sky density.  In principle, a fraction of sources 
identified as stars or ``normal'' galaxies (or elusive AGN, see discussion
in Sec.~7.5) could be spurious counterparts. 
Considering the density of stars and 
galaxies at the faintest magnitudes observed in the two classes of sources
(R$_{stars}\leq$18 and R$_{galaxies}\leq$21) we expect about 12 stars and
4 galaxies falling by chance within 4$\arcsec$ from the 400 X-ray positions.
This is clearly an upper limit given the adopted identification process: 
we have usually observed all the bright (i.e. visible on the DSS) objects 
falling within the circle of 4$\arcsec$ radius and, whenever an AGN is found 
we have 
considered it as the right counterpart (as described above the probability
of finding an AGN by chance is very low in our survey) and discarted 
the others (either stars or galaxies). This strategy excludes the large 
majority of possible spurious galaxy or star identifications: only those
stars or galaxies falling by chance close to an X-ray source whose real 
counterpart is weak (e.g weaker than the
DSS limit) have the possibility of being considered the counterpart by
mistake. Since the majority ($\sim$90\%) of the real counterparts are 
expected to be brighter than the DSS limit, we conclude that only $\sim$1/10
of the 12 stars and 4 galaxies falling by chance in the error circle
have the possibility of being considered as the counterpart. Therefore, the
actual number of spurious stars and galaxies in the sample should  be 
$\sim$1.2 and $\sim$0.4 respectively. In conclusion, we do not expect
more than 1-2 mis-identifications in the entire XBS survey.

\section{Optical spectroscopy}
About 2/3  of the spectroscopic identifications (i.e. $\sim$240 objects) 
of the XBS survey come
from dedicated spectroscopy carried out during 5 years (from 2001 to 2006) 
at several optical telescopes. Most of the identifications are obtained at 
the Italian Telescopio Nazionale Galileo (TNG, 51\% of the 
identifications) and at the ESO 3.6m and NTT telescopes (37\%). The 
remaining 12\% has been collected from other telescopes like the
 88" telescope of the University of Hawaii (UH) in Mauna Kea and the 
Calar Alto 2.2m telescope.

The instrumental configurations are summarized in
Table~\ref{setup}. 
We have always adopted a long-slit configuration with  
low/medium dispersion  (from 1.4 \AA/pixel
to 3.7 \AA/pixel) and low/medium resolution (from $\sim$250 to 450) 
grisms to maximize the wavelength coverage.
For the data reduction we have used the IRAF {\it longslit} package. 
The spectra have been wavelength calibrated using a reference spectrum and 
flux calibrated using photometric standard stars observed during the
same night. Most of the observations were carried out during non-photometric
conditions. Since the main goal of the observations was to secure a
redshift and a spectral classification of the source we did not attempt
to obtain an absolute flux calibration of the spectra.  

\begin{table*}
\begin{center}
\begin{tabular}{l l l c}
Telescope/instrument/grism & slit width & dispersion 
& Observing nights \\
\ & (arcsec) & (\AA/pixel) & \\
\ & \ & \ &  \\
\hline
 TNG+DOLORES+LRB      &   1.5, 2    &          2.8 &   15-16/11/2001        \\   
 UH88"+WFGS+green(400)  &   1.6       &          3.7 &   16-18/04/2002        \\   
 ESO3.6m+EFOSC+Gr6      &   1.2       &          2.1 &   02-08/05/2002        \\   
 TNG+DOLORES+LRB      &   1.5       &          2.8 &   23/06/2002           \\   
 TNG+DOLORES+LRB      &   1.5       &          2.8 &   09-12/09/2002        \\   
 ESO3.6m+EFOSC+Gr13     &   1.2, 1.5  &          2.8 &   30/09-02/10/2002     \\  
  TNG+DOLORES+LRB      &   1.5       &          2.8 &   05-07/10/2002        \\ 
 CA2.2m+CAFOS+B200/R200 &   1.5       &     4.7/4.3 &   30/10-01/11/2002     \\   
 TNG+DOLORES+LRB      &   1.5       &          2.8 &   25-27-28/12/2002     \\   
 TNG+DOLORES+LRB      &   1.5       &          2.8 &   27-30/03/2003        \\  
  NTT+EMMI+Gr3        &   1.0         &          1.4 &   02-03/05/2003        \\ 
 TNG+DOLORES+LRB      &   1.5       &          2.8 &   08/05/2003           \\   
  TNG+DOLORES+LRB      &   1.5       &          2.8 &   27/09-01/10/2003     \\   
  NTT+EMMI+Gr2        &   1.0, 1.5    &          1.7 &   04-06/01/2005        \\   
 TNG+DOLORES+LRB      &   1.5       &          2.8 &   12-16/03/2005        \\   
 NTT+EMMI+Gr2        &   1.5       &          1.7 &   07/10/2005           \\   
 NTT+EMMI+Gr2        &   1.0, 1.5, 2.0 &          1.7 &   02-05/03/2006        \\   

\hline
\end{tabular}
\end{center}
\caption{Journal of observations}
\label{setup}

\vspace*{0.3 cm}

%$^{a}$ {in arcseconds}

%$^{b}$ {in \AA/pixel}

\end{table*}

In general, we have two exposures for each object, except for
a few cases in which we have only one spectrum or three exposures. 
Cosmic rays were subtracted manually from the extracted spectrum or 
automatically if three exposures of equal length are available. 

On average, the seeing during the observing runs ranged from  
1\arcsec\ to 2\arcsec\ with few exceptional cases of seeing below
1\arcsec\ (0.5\arcsec-0.8\arcsec, typically during the runs at the ESO NTT).
Usually, during very bad seeing conditions ($\geq$2.5\arcsec) no
observations have been carried out.
We have typically used  a slit width of 1.2\arcsec-1.5\arcsec\ except for the periods
of sub-arcsec seeing conditions, where a slit width of 1\arcsec\ was used 
to maximize the signal. 

\section{Data from the literature}
The remaining 1/3 of the spectroscopic identifications of the XBS survey 
have been taken from the literature (NED and SIMBAD\footnote{NED (NASA/IPAC Extragalactic 
Database) is operated by the Jet Propulsion Laboratory,
California Institute of Technology, under contract with the
National Aeronautics and Space Administration; SIMBAD is operated at CDS 
(Strasbourgh, France).}) or from other XMM-Newton 
identification programs like AXIS (Barcons et al. 2007). 
Whenever possible we have
obtained the optical spectrum of the extragalactic sources, either in 
FITS format or a printed spectrum, and then
analysed it using the same criteria adopted for the spectra collected
during our own observing runs. In few cases we have not found a spectrum but 
tables
presenting the relevant pieces of information on the emission lines. 
Therefore the spectral analysis (for classification purpose) 
has been possible for nearly all 
the extragalactic identifications coming from the literature or from
the AXIS program. 

If a classification is present in the literature but no further information
is found we have kept the classification only if it can be considered
unambiguous (e.g. a type~1 QSO,  see discussion in Section~7). 

\section{Spectral analysis}
For more than 80\% of the extragalactic identifications (either from our own 
spectroscopy or from the literature) we have an optical spectrum 
in electronic format. We have used the task ``splot'' within the $iraf$
package to analyse these spectra and get the basic pieces of information,
like the line positions, equivalent widths (EW) and FWHM. 
During the fit we use a Gaussian or a  Lorentzian profile. When two components
are clearly present in the line profile (e.g. a narrow core plus a broad 
wing) we attempt a de-blending. 

Given the moderate resolution of the spectroscopic observations 
(FWHM$\sim$650-1200 km s$^{-1}$) we have applied a correction to the
line widths to account for the instrumental broadening, i.e.:

\begin{center}
$\Delta\lambda = \sqrt{\Delta\lambda_o^2 - \Delta\lambda_{inst}^2}$
\end{center}

where $\Delta\lambda$, $\Delta\lambda_o$ and $\Delta\lambda_{inst}$ are the 
intrinsic, the observed and the instrumental line width 
respectively. 

The errors on 
EW and FWHM have been estimated with the task ``splot''. This task adopts 
a  model  for the  pixel  sigmas based on a Poisson statistics model of the 
data.  
The model parameters are  a  constant Gaussian  sigma and an "inverse gain".
We have set this last parameter to ``0''  i.e. we assume that the part of 
noise due to 
instrumental effects (RON) is negligible. This is reasonable for 
our spectra. 
The  de-blending  and  profile  fit  error  estimates are computed by
Monte-Carlo simulation (see {\it iraf} help for details).  
We found that the errors computed in this way are sometimes underestimated, in 
particular when the background around the emission/absorption line is
not well determined and/or when the adopted model profile (Gaussian or
Lorentz profile) does not give a correct description of the line. In these 
cases we have adopted a larger error that includes the values obtained with
different background/line profile models. 

For all the identifications for which only a printed spectrum is
available, we 
have performed a similar (but rougher) analysis and included the 
larger uncertainties in the error bars.

\section{Spectroscopic classification and redshift}
On the basis of the data collected from the literature and the spectra 
obtained from our own spectroscopy, we have determined a spectroscopic 
classification and a redshift  for 87\% (348) of the XBS objects.
The sources can be broadly grouped into stars, clusters of galaxies and 
AGN/galaxies. Stars and AGN/galaxies represent the most numerous populations
in the sample, being 17\% and 80\% respectively of the total number
of the identified XBS sources. An extended  analysis of the X-ray and optical 
properties of the 58 stars found in the sample has been
already presented in L\'opez-Santiago et al. (2006) and will not
be discussed in this paper anymore. 

The classification of a XBS source as a cluster of galaxies is essentially
based on the visual detection of an over density of sources in the 
proximity of the X-ray position on the optical image and on the spectroscopic 
confirmation that some of these
objects have the same redshift. In all these cases, the object closer to the
X-ray position is an optical ``dull'' elliptical galaxy. 
The cluster nature of the XBS sources
is usually confirmed by a visual inspection of the X-ray
image which  shows that the X-ray source is extended.
In the XBS survey we currently have only 8 objects classified as  
clusters of galaxies. However, this type of objects is certainly 
under-represented because the source detection algorithm is optimized
for point-like sources (see Della Ceca et al. 2004). This is true also
for normal galaxies whose X-ray emission (due to diffuse gas and/or 
discrete sources) is extended.

In this paper we will not discuss any further stars and clusters of galaxies 
and they will be excluded from the following analysis. 
In this section we present in  detail the 
criteria  adopted to classify the remaining extragalactic sources, i.e. AGNs
(including BL Lac objects) and galaxies.

%                                     Two column figure (place early!)
%______________________________________________ Gamma_1 (lg rho, lg e)
   \begin{figure*}
   \centering
   \includegraphics[width=16cm,height=16cm]{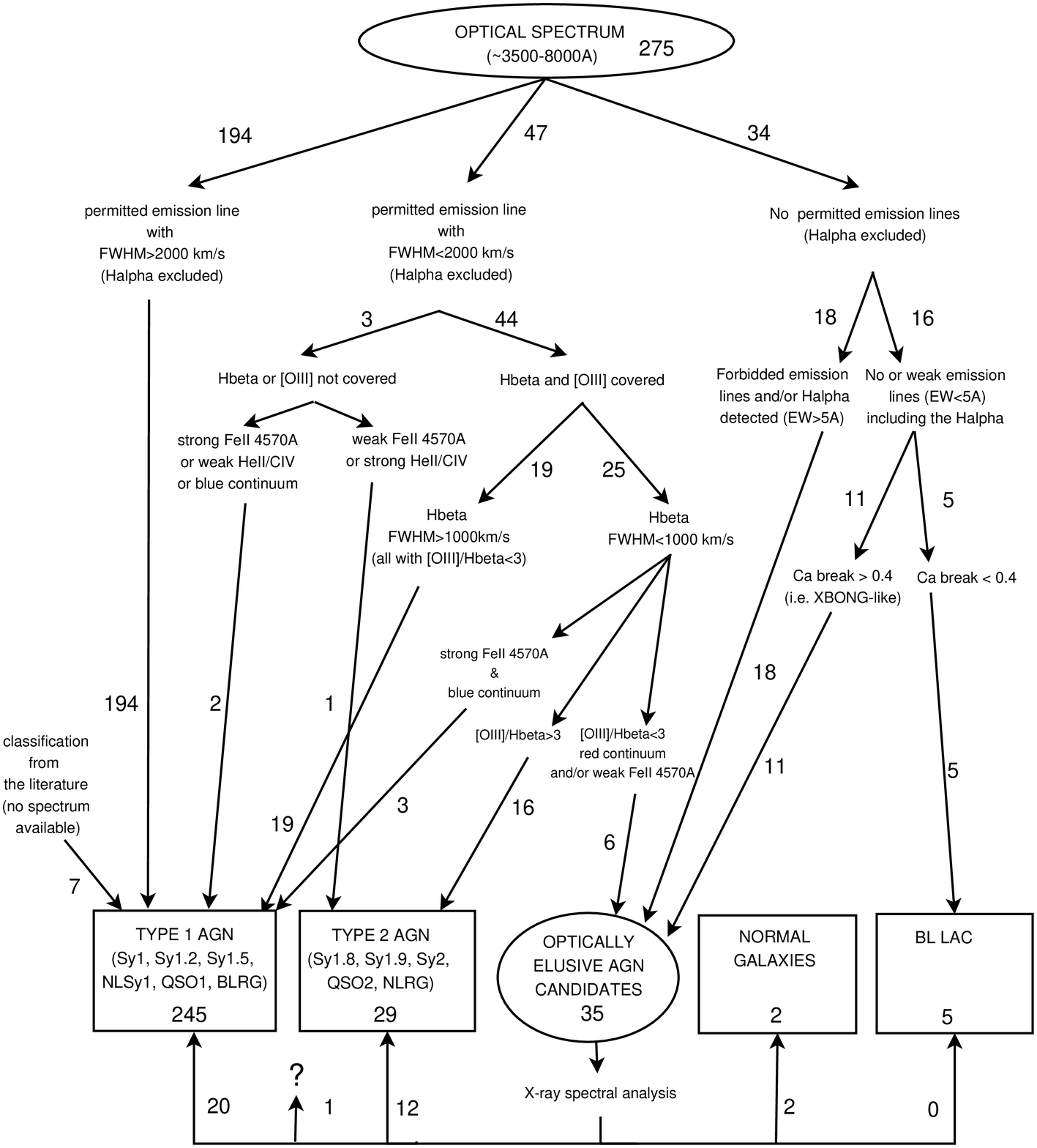}
   \caption{Classification flow-chart of the XBS extragalactic sources 
(excluding clusters of galaxies). Numbers near the arrows indicate the
number of XBS sources that have followed the correspondent path. The sources 
within the ``optically elusive AGN candidates'' group have been
classified on the basis of the X-ray spectrum and re-distributed into the 
other classes accordingly (see Caccianiga et al. 2007). 
%for which
%the H$\beta$/[OIII]$\lambda$5007\AA\ spectral region is not covered 
%(z larger than $\sim$0.65).
}
              \label{classification}%
    \end{figure*}

 \subsection{The classification scheme}
The large majority ($\sim$90\%) of the extragalactic sources in the XBS survey
show strong (EW$>$10 \AA) emission lines in the optical spectrum. In most 
of these objects the analysis of the emission lines gives a clear
indication of the presence of an AGN. 

One of the  primary goals of the XBS 
survey is to explore the population of 
absorbed AGN. For this reason, we want to adopt an optical 
classification that can reliably separate optically absorbed 
from non-absorbed objects. 
The criterion typically used to separate optically absorbed and non-absorbed
AGN is based on the width of the permitted/semi-forbidden emission lines, 
when present. 
However, different thresholds have been used in the literature
to distinguish type~1 (i.e. AGN with broad permitted or semi-forbidden 
emission lines) and
type~2 AGN (i.e. with narrow permitted/semi-forbidden emission lines) 
ranging from 1000 km s$^{-1}$ (e.g. Stocke et al. 1991 for the
Extended Medium Sensitivity Survey, EMSS) up to 
2000 km s$^{-1}$ (e.g. Fiore et al. 2003, for the Hellas2XMM survey). 
Both thresholds present some limits.

From the one hand, the 2000 km s$^{-1}$ threshold may mis-classify 
the Narrow Line Seyfert 1 (NLSy1) and their high-z
counterparts, the ``Narrow Line QSO'' (NLQSO, see for instance Baldwin 
et al. 1988), as type~2 AGN.
These sources typically show
permitted/semi-forbidden lines of width between 1000 and 2000 km s$^{-1}$ 
(or even lower, 
see for instance V\'eron-Cetty, V\'eron \&  Gon{\c c}alves2001) but 
it is generally 
accepted that the relatively narrow permitted/semi-forbidden lines
are not due to the presence of strong optical absorption but, rather, 
they are
connected to the physical conditions of the nucleus (e.g. Ryan et al. 2007 and
references therein).

On the other hand, the adoption of a lower threshold (e.g. 1000 km s$^{-1}$) 
to distinguish type~1 and type~2 AGNs can systematically mis-classify
high-z QSO 2, where the observed permitted lines are typically between
1000 and 2000 km s$^{-1}$ (e.g. Stern et al. 2002; Norman et al. 2002; 
Severgnini et al. 2006).

%                                     One column figure (place early!)
   \begin{figure}
   \centering
    \includegraphics[width=9cm]{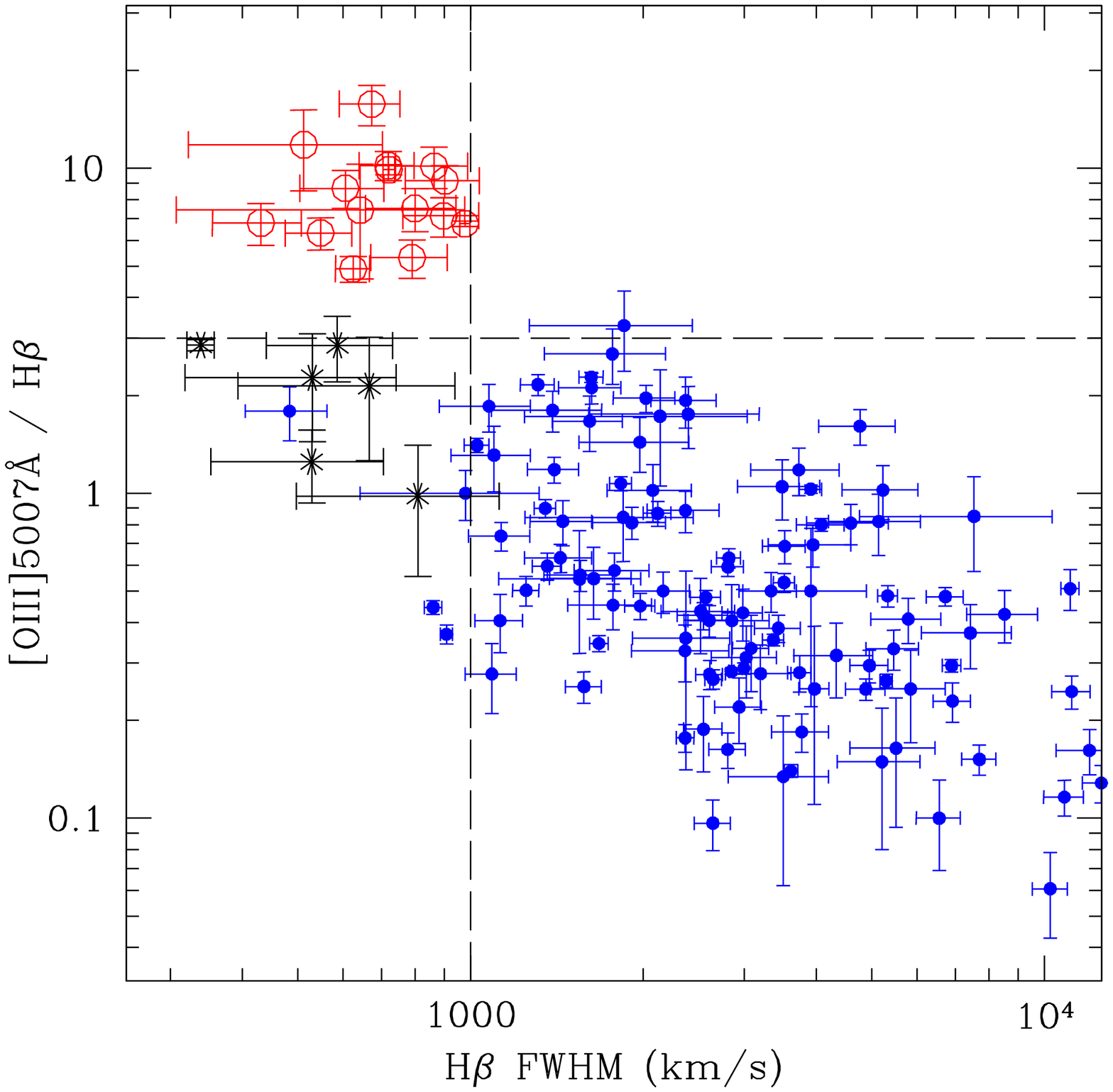}

   \caption{[OIII]$\lambda$5007\AA\ /H$\beta$ flux ratio versus the
H$\beta$ emission line width for the sources in the XBS survey for which
these lines have been detected 
(z below $\sim$0.65). Filled points are type~1 AGN, open circles are
type~2 AGN, stars are objects spectroscopically classified as 
emission line galaxies (HII-region/starburst galaxies).
The two dashed lines indicate the reference values used for the
spectroscopic classification (see text and Fig.~\ref{classification}
for details)
}
              \label{diag2}
    \end{figure}

It is thus clear that a simple classification exclusively based on the
widths of the permitted lines cannot be realistically adopted. Additional
diagnostics are necessary for a reliable optical classification. 

In Figure~\ref{classification} we present the flow-chart that summarizes
the classification criteria used for the XBS extragalactic sources 
(excluding the clusters of galaxies). The complexity of the 
presented flow-chart is mainly due to the fact of dealing with
sources distributed along a wide range of redshift (from local
objects up to z$\sim$2): the emission lines that can be used
for the spectral classification are thus different depending on
the redshift of the source.
Another source of complexity is the
problem of optical ``dilution'' due to the host-galaxy light (see below).

The final classes (represented by 4 boxes) are type~1 AGN, 
type~2 AGN, BL Lac objects and the ``normal'' (i.e. not powered by an AGN) 
galaxies. In 35 cases the optical spectrum is dominated by
the star-light from the host-galaxy and establishing the presence of an AGN 
and its type (e.g. type~1 or type~2) through the optical spectrum is not
possible. For this group of objects, named ``optically elusive AGN'' 
candidates, we have used the
X-ray data to asses the presence of an AGN and to characterize its
nature (i.e. absorbed or unabsorbed, see Caccianiga et al. 2007 and 
Section~7.5 for details). 

We have 
considered in the type~1 AGN class the intermediate types 1.2 and 1.5,
while the type~2 AGN class includes the 1.8 and 1.9 types.
This distinction is expected to correspond to a separation into a level of 
absorption
lower/larger than A$_V\sim$2 mag (see discussion below), i.e. a column density 
(N$_H$) 
larger/lower than $\sim$4$\times$10$^{21}$ cm$^{-2}$ assuming a 
Galactic standard N$_H$/A$_V$ conversion. 

We have applied these steps to the
275 objects for which the required information is available (either
from our own spectroscopy or from the literature). Besides these
275 we have 7 additional objects whose classification has been 
taken from the literature but it is not possible to directly
apply the classification criteria discussed here since a spectrum or
a table reporting the lines properties is not available. These 7 objects
are all classified as type~1 AGN with redshift between 0.64 and 1.4 and
X-ray luminosities between 10$^{44}$ and 10$^{46}$ erg s$^{-1}$ (i.e.
they are type1 QSO). We have adopted  the published classification 
for these objects even if they have not passed through the 
classification steps presented in Fig.~\ref{classification}.

We briefly discuss here the main classification steps presented in
Fig.~\ref{classification}. 

\subsection{AGN with broad (FWHM$>$2000 km s$^{-1}$) permitted 
emission lines}
The first main ``arrow'' of Figure~\ref{classification} 
considers the detection of one (or more) 
very broad (FWHM$>$2000 km s$^{-1}$) permitted emission line. 
In this step we do not consider the H$\alpha$ line. 
The reason is that, whenever only a strong and broad H$\alpha$ emission line
is detected in the optical spectrum it is not possible to correctly
classify the object. Indeed, sources where only a broad H$\alpha$ line 
is clearly 
detected can be both unabsorbed AGN or intermediate AGNs, like Sy1.8 or 
Sy1.9. Since, as discussed above, we consider Sy1.8 and Sy1.9 as type~2 AGN 
we are not able to correctly classify these sources as type~1 or type~2 just 
on the basis of the H$\alpha$ line.

This first step allows to directly classify as type~1 AGN all the sources
with very broad (FWHM$>$2000 km s$^{-1}$) permitted/semi-forbidden emission lines.
These sources are ``classical'' type~1 AGN (Sy1 and QSO).

\subsection{Objects with permitted emission lines with FWHM$<$2000 km s$^{-1}$}

The second main ``arrow'' regards  sources for which 
``narrow'' (FWHM$<$2000 km s$^{-1}$) permitted emission lines (H$\alpha$ excluded) are detected. 
As discussed above, in this group many
different types of sources can be found, including absorbed AGNs, 
AGNs with intrinsically narrow permitted/semi-forbidden emission lines 
(NLSy1 and NLQSO) and emission-line galaxies like starburst/HII-region 
galaxies.
As already stressed, a proper classification of these objects requires
the application of diagnostic criteria. For the sources at relatively low z 
(below $\sim$0.65) the detection of two
critical emission lines, i.e. the H$\beta$ and the [OIII]$\lambda$5007\AA\,
can significantly help the classification. We thus discuss separately 
the sources according to the fact that the H$\beta$/[OIII]$\lambda$5007\AA\
spectral region is covered (i.e. sources with z below $\sim$0.65) or
not (i.e. sources with z larger than $\sim$0.65).

{\bf H$\beta$ and [OIII]$\lambda$5007\AA\ covered.}

In all but 3 objects with strong and relatively narrow (FWHM$<$2000 km s$^{-1}$) 
permitted/semi-forbidden emission lines  the  H$\beta$ and 
[OIII]$\lambda$5007\AA\ spectral region is covered.
As discussed by several authors 
(e.g. Veron-Cetty \& Veron 2003; Winkler 1992; Whittle 1992), a clear
distinction between different types of  AGN can be based on the 
ratio between [OIII]$\lambda$5007\AA\ and H$\beta$ line intensity.
Optically absorbed Seyferts, like Seyfert~2 or Seyfert1.8/1.9, present
high values of the  [OIII]$\lambda$5007\AA\ /H$\beta$ flux ratios ($>$3),
while moderately absorbed or non-absorbed Seyferts (Seyfert~1.5, Seyfert~1.2
and Seyfert 1 and NLSy1) show a  [OIII]$\lambda$5007\AA\ /H$\beta$ flux ratio 
between 0.2 and 3. In Figure~\ref{diag2} we show the 
[OIII]$\lambda$5007\AA\ /H$\beta$ flux ratio versus the H$\beta$ width
for all the XBS sources for which these lines are observed (including 
sources with FWHM$>$2000 km s$^{-1}$ emission lines). The two
quantities are strongly coupled, and the objects with broad ($>$1000 km s$^{-1}$)
H$\beta$  have all (but one) [OIII]$\lambda$5007\AA\ /H$\beta$ 
flux ratio below 3. 
We classify all these objects type~1 AGN, including the source for which
the  [OIII]$\lambda$5007\AA\ /H$\beta$ flux ratio is marginally greater
than 3, since the value is consistent, within the errors, with those observed 
in type~1 AGN.

On the contrary,
the objects with a narrow ($<$1000 km s$^{-1}$) H$\beta$ present a wide range
of  [OIII]$\lambda$5007\AA\ /H$\beta$ flux ratios, from 0.3 to 15. This latter
class of sources includes both type~2 AGN, ``normal'' galaxies 
(e.g. HII-region galaxies or starburst galaxies) and some NLSy1. 
To distinguish all
these cases it is necessary to apply the diagnostic criteria discussed,
e.g. in Veilleux \& Osterbrock 1987, to separate type~2 AGN from
HII-region/starburst galaxies, and/or the diagnostics based, for instance
on the FeII$\lambda$4570\AA\ /H$\beta$ flux ratio to recognize the
NLSy1 (V\'eron-Cetty, V\'eron \& Gon{\c c}alves 2001). The adopted criteria  
are indicated near the correspondent arrows of Fig.~\ref{classification}.

{\bf H$\beta$ and/or [OIII]$\lambda$5007\AA\ not covered.}

Only for 3 sources with strong and relatively narrow (FWHM$<$2000 km s$^{-1}$)
permitted emission lines the H$\beta$/[OIII]$\lambda$5007\AA\ spectral range
is not covered. 
As discussed above these sources can be 
optically absorbed AGN (i.e. type~2 AGN) or NLQSO. The distinction between
these two classes at large redshift is more critical than at lower z and other
diagnostics must be used, like 
the intensity of the FeII$\lambda$4570\AA\ hump or the 
relative strength of the HeII emission line when compared to the
CIV$\lambda$1549\AA\ (e.g. Heckman et al. 1995).

One of these objects (XBSJ021642.3--043553, z=1.985) 
has been extensively discussed in
Severgnini et al. (2006) and it is classified as type~2 QSO on the basis
of the relative strength of the HeII emission line when compared to the
CIV$\lambda$1549\AA.

The second source (XBSJ120359.1+443715, z=0.541) has a blue spectrum and a 
quite strong Fe~II4570\AA\ hump which is usually considered as the signature 
of 
a NLSy1. Unfortunately we cannot further quantify the strength of this hump
in respect to the H$\beta$ line since this latter line falls outside
the observed spectrum. We classify this object as NLQSO candidate. 

Finally, in the third object (XBSJ124214.1--112512, z=0.82) we have detected 
the MgII$\lambda$2798\AA\ emission line with a relatively narrow 
(FWHM$\sim$1900 km s$^{-1}$) core plus a broad wing. Both the FeII$\lambda$4570\AA\ 
and the HeII line fall outside the observed range and we cannot apply the 
diagnostic criteria discussed above. 
Using the spectral model described in Section~8 
we have successfully fitted the observed continuum emission using
a value of A$_V\sim$0.5 mag i.e. below the 2 mag limit that corresponds to our
classification criteria (see Section~8). We thus classify this object as 
type~1 AGN. 

\subsection{Sources with weak (or absent) permitted emission lines}
The last main ``arrow'' of Figure~\ref{classification} corresponds to
sources with no (or weak) permitted emission lines (excluding the
H$\alpha$ line, as discussed above). 
This group of sources includes both ``featureless'' AGN (the BL Lac objects)
and sources whose optical spectrum is dominated by the host-galaxy and
no evidence (or little evidence) for the presence of an AGN can be 
inferred from the optical spectrum. As already discussed, these
latter objects are considered as ``elusive'' AGN candidates and 
analysed separately using the X-ray information (see next sub-section).

BL Lac objects are classified on the basis of the lack of
any (including the H$\alpha$) emission line and the shape of
the continuum around the 4000\AA\ break 
($\Delta$\footnote{The 4000\AA\ break is
defined as $\Delta$ = $\frac{F^+ - F^-}{F^+}$ where F$^+$ and F$^-$ 
represent the mean value of the flux density 
(expressed per unit  frequency) in the region 4050 - 4250~\AA\ and
3750 - 3950~\AA\ (in the source's rest-frame) respectively.}
).  
In fact the detection of  a significant
reduction of the 4000\AA\ break  when compared with elliptical 
galaxies is considered as an indication for the presence of nuclear 
emission.  We adopt the limit commonly used in the literature of $\Delta<$40\%
to classify the source (with no-emission lines) as a BL Lac object 
(e.g. see the discussion in Landt et al. 2002).

The BL Lacs are 5 in total and all have been
detected as radio sources in the NVSS 
(Condon et al. 1998) radio survey, something which is considered as a further
confirmation of the correct classification. The properties of the XBS  
BL Lacs are presented in Galbiati et al. (2005). 
As discussed in Caccianiga et al. (2007) we cannot exclude that some of the
``elusive'' AGN  are actually hiding a BL Lac nucleus.
The best way to find them out is through a deep radio follow-up. On the
basis of the current best estimate of the BL Lac sky density, however, we
do not expect more than 1-2 BL Lacs hidden among the XBS elusive AGN.

 \subsection{The optically ``elusive'' AGN candidates}
As summarized in Fig.~\ref{classification}, different classification paths
lead to the group of optically ``elusive'' AGN candidates. 
All these sources (35 in total) are characterized by the presence, 
in the optical spectrum,
of a significant/dominant contamination of star-light from the
host galaxy. In some cases, i.e. for the so-called  X-ray Bright
Optically Normal Galaxies (XBONG) and the HII-region/starburst 
galaxies, we do not have clear (optical) evidence 
for the presence of an AGN. 
We consider as optically ``elusive'' AGN candidates also the 
sources where a broad 
($>$1000-2000 km s$^{-1}$) H$\alpha$ line is probably present 
but where most of the remaining emission lines (in particular
the H$\beta$ emission line) are not detected. Even if the presence of 
an AGN in these sources
is somehow suggested by the detection of a  broad H$\alpha$ emission line, the 
``dilution'' due  to the host galaxy
is critical also in these cases because it does not permit a quantification 
of the optical absorption (i.e. type~1 or type~2 AGN).
Similarly, some other sources in this group show a quite strong 
[OIII]$\lambda$5007\AA\ , which can be suggestive of the presence
of an AGN, but no H$\beta$ is detected, something that prevents us from
a firm  classification of the source.

Given the objective difficulty of using the optical spectra to assess 
the actual presence of an AGN and to give a correct classification of  it
(i.e. type~1, type~2 or BL Lac object) we have analyzed the X-ray data.
In particular, we have shown that the X-ray 
spectral shape combined with the X-ray luminosity of the sources 
allows us to assess 
the presence of an AGN and to quantify its properties.
While the detailed discussion of this analysis has been reported in
Caccianiga et al. (2007) we summarize here the main conclusions.
In the large majority of cases (33 out of 35 objects) the X-ray analysis 
has revealed an AGN while only in 2 cases the X-ray emission is probably
due to the galaxy (either due to hot gas or to discrete sources) 
given the low X-ray luminosities (10$^{39}$-10$^{40}$ erg s$^{-1}$). 
In 20 sources where an AGN has been detected the column densities 
observed are below N$_H$=4$\times$10$^{21}$ cm$^{-2}$ while in 12 
the values are higher. Only for one object the data do not allow the
estimate of the column density. According to the Galactic relationship
between optical (A$_V$) and X-ray absorption (N$_H$) the 
value of N$_H$=4$\times$10$^{21}$ cm$^{-2}$ corresponds to A$_V\sim$2 mag
which is the expected dividing line between type~1 and type~2 sources 
as defined in this paper, i.e. following the scheme of 
Fig.~\ref{classification} (see the discussion in Section~8).
We have thus classified these 32 ``elusive'' AGN into type~1 and type~2
according to the value of N$_H$ measured from the X-ray analysis. 
In Tab~\ref{table} these classifications are flagged to
indicate that they are not based on the optical spectra. 

\section{Diagnostic plots}
Using a simple spectral model, discussed in Severgnini et al. (2003), we
have produced some diagnostic plots that may help in the
classification of X-ray selected sources. 
This model  
uses an AGN template  composed of two parts: a) 
the continuum with the broad emission lines and b) the narrow emission lines.
According to the basic version of the AGN unified model, the first part can 
be absorbed while the second one is not affected by the presence of
an obscuring medium. 
The AGN template is based on the data taken 
from Francis et al. (1991) and Elvis et  al. (1994) while the extinction curve 
is taken from Cardelli, Clayton \& Mathis (1989). 
Besides the AGN  template, 
the spectral model includes also a galaxy template, produced on the basis of 
the Bruzual \& Charlot (2003) models. 

We have then applied different levels of A$_V$ and measured the expected values
of some critical quantities like 
the 4000\AA\ break, the [OIII]$\lambda$5007\AA\ 
and the H$\alpha$ line equivalent width. 

  \begin{figure}
   \centering
    \includegraphics[width=9cm]{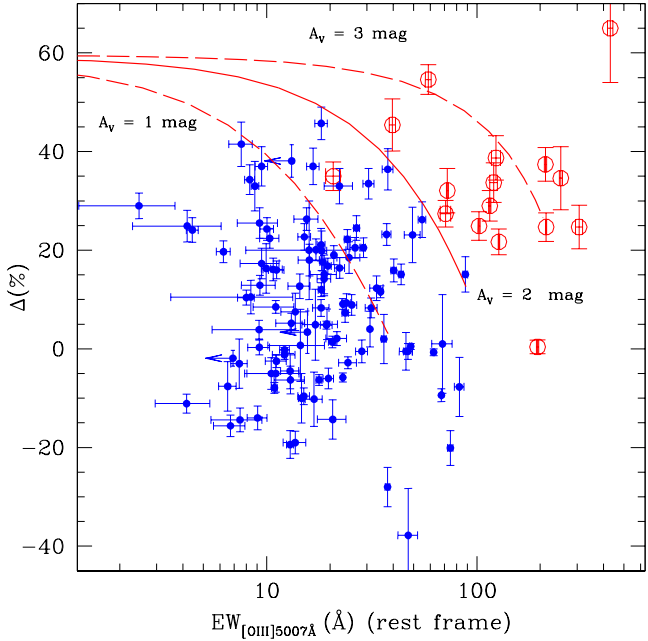}

   \caption{4000\AA\ break ($\Delta$) versus the [OIII]$\lambda$5007\AA\ 
equivalent widths for the
BSS type~1 (filled points) and type~2 (open circles)
AGNs, excluding the elusive ones. 
The three lines show the expected 
regions corresponding to
different optical absorptions, from A$_V$=1 mag to A$_V$=3 mag, assuming a 
10 Gyr early-type host galaxy}
              \label{loiii_c_paper_10G}
    \end{figure}

 \begin{figure}
   \centering
    \includegraphics[width=9cm]{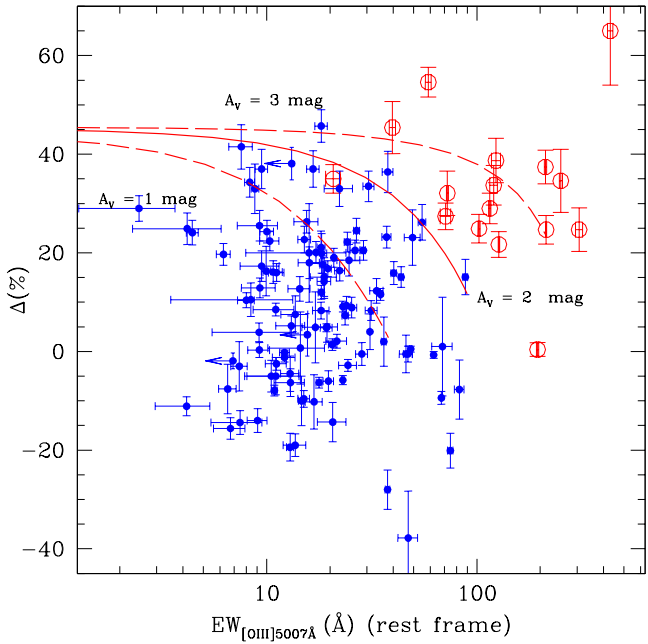}

   \caption{Same plot as in fig.~\ref{loiii_c_paper_10G} but assuming a 
younger (1 Gyr) host-galaxy. Symbols as in Fig. 6.}
              \label{loiii_c_paper_1G}
    \end{figure}

\subsection{Non-elusive AGN}
In Fig.~\ref{loiii_c_paper_10G} we have plotted the 4000\AA\ break versus the
[OIII]$\lambda$5007\AA\ equivalent width of all the XBS sources classified 
as type~1 o type~2 AGN and for which these
quantities have been computed, excluding the elusive AGN. 
On this plot type~2 and type~1 AGN occupy separated regions, with type~2 AGN 
showing the largest [OIII]$\lambda$5007\AA\ equivalent widths and largest 
4000\AA\ breaks. This separation is expected since the presence of a large level 
of absorption in these sources significantly suppresses the AGN continuum,
from the one hand, and increases the narrow lines equivalent widths, on the
other hand. 
On the same figure we have then plotted the curves based on the spectral model
described above for three different values of absorption, from A$_V$=1 mag to 3 mag,
assuming a 10 Gyr old early-type host galaxy. The A$_V$=2 mag curve is clearly the
one that better separates the two classes of AGNs. This result does not depend significantly
on the host-galaxy type as shown in Fig.~\ref{loiii_c_paper_1G}, where a much younger 
host-galaxy
is assumed (1 Gyr). Also in this plot the line that better separates type~1 and type~2 AGN
is the one corresponding to A$_V$=2 mag. This weak dependence with the host-galaxy type
is not anymore true if we consider the elusive AGN i.e. those sources whose optical
spectrum is dominated by the host-galaxy and that occupy the upper-left region of the
diagram. Therefore, this plot cannot be used as diagnostic for the elusive AGN.

\begin{figure}
   \centering
    \includegraphics[width=9cm]{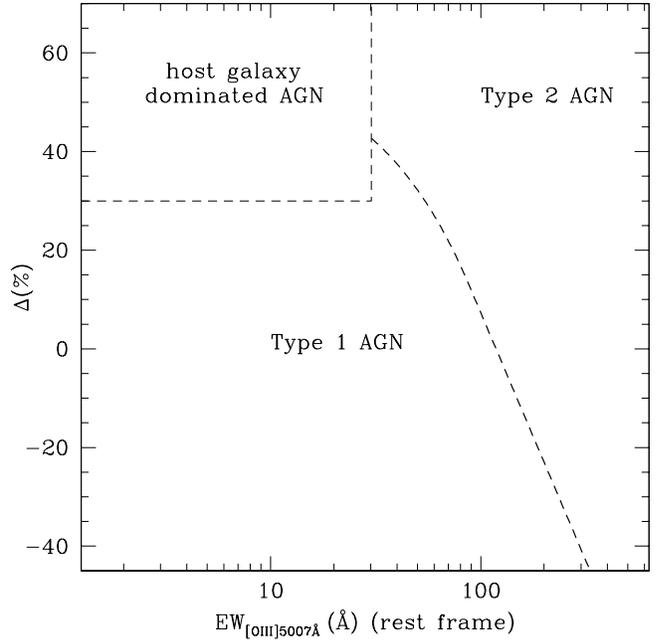}

   \caption{Typical regions occupied by the XBS AGNs on the 4000\AA\ break/[OIII]$\lambda$5007\AA\ 
EW plot}
              \label{regions2}
    \end{figure}

The clear separation between type~1 and type~2 AGN observed in 
a [OIII]$\lambda$5007\AA/4000\AA\ plot can be used as a simple diagnostic, at least for
objects not dominated by the host-galaxy light. In Fig.~\ref{regions2} we report the typical
regions occupied by type~1 and type~2 ANGs and (most of) the elusive AGN.
This diagnostic diagram is simple
to apply, requiring just the measure of the fluxes across the 4000\AA\ break and the
[OIII]$\lambda$5007\AA\ equivalent width,  and can be used up to z$\sim$0.8 
(or higher if infrared spectra are available).

\subsection{Elusive AGN with a broad H$\alpha$ emission line}
By definition, elusive AGN have an optical spectrum which is dominated
by the host galaxy light and, therefore, it is difficult/impossible 
to obtain a clear classification directly from the optical data. 
However, as already mentioned, 
in a number of elusive AGNs  a possibly broad H$\alpha$ line 
in emission is found. 
In itself, this piece of information cannot give a clear indication
of the type of AGN present in the source. 
With the support of the spectral model previously discussed 
we now want to find a method to estimate the level of
optical absorption in these sources. We want to use  only the few AGN emission 
lines that usually can be detected
even in the presence of a high level of dilution, i.e. the 
[OIII]$\lambda$5007\AA\ and the H$\alpha$ emission 
lines. 

%                                     One column figure (place early!)
%______________________________________________ Gamma_1 (lg rho, lg e)
   \begin{figure}
   \centering
    \includegraphics[width=9cm]{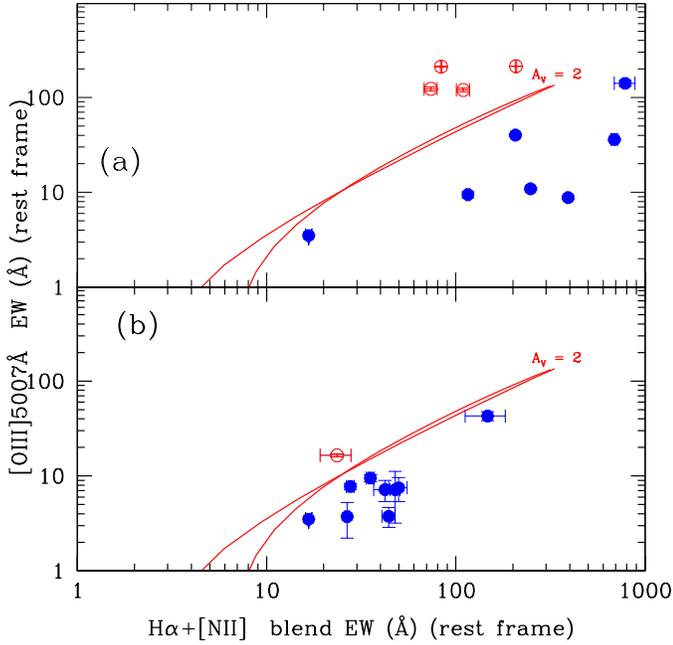}

   \caption{[OIII]$\lambda$5007\AA\  versus the H$\alpha$+[NII] line blend 
equivalent widths of the XBS AGN classified as type~1 and type~2 on the
basis of the optical spectrum (panel a) 
and of the elusive AGN for which a broad H$\alpha$ emission line 
has been detected (panel b).
In this case the classification is based on the X-ray spectrum. Open
circles are type~2 AGN  while filled points are type~1 AGN.
Solid lines
show the theoretical separation between objects with large  ($A_V>$2 mag) 
and small ($A_V>$2 mag) optical absorption corresponding to a threshold of
N$_H$=4$\times$10$^{21}$ cm$^{-2}$ assuming a Galactic standard relation.
The two lines correspond to different ages of the host-galaxy (t=1 Gyr and 
10 Gyr respectively).
}
              \label{oiii_ha}
    \end{figure}

Interestingly, the combination of the H$\alpha$ line intensity with 
the [OIII]$\lambda$5007\AA\ emission line can help the classification of the
source. 
In Figure~\ref{oiii_ha} we show the [OIII]$\lambda$5007\AA\  versus the 
H$\alpha$+[NII] blend\footnote{The reason for using the blend instead of 
the single H$\alpha$ line is that, in most cases, the three lines (H$\alpha$, 
[NII]$\lambda$6548\AA, [NII]$\lambda$6583\AA) are blended together
and it is not easy (or possible) to disentangle the different contributions} 
equivalent widths of the XBS AGN classified as type~1 and type~2 on the
basis of the optical spectrum (panel~a). In panel (b) we report the
9 elusive AGN with a broad (FWHM$>$1000 km s$^{-1}$) H$\alpha$ emission line. 
In this
case, the symbols represent a classification based on the X-ray 
spectral analysis, i.e. open
symbols are AGN with N$_H>$4$\times$10$^{21}$ cm$^{-2}$ while filled
circles are AGN with  N$_H<$4$\times$10$^{21}$ cm$^{-2}$. 
On the two panels we report also the theoretical lines that separate
between AGNs with large  ($A_V>$2 mag) 
and small ($A_V<$2 mag) optical absorption (corresponding to 
N$_H$ larger or lower than 4$\times$10$^{21}$ cm$^{-2}$ assuming a 
Galactic standard relation). 
Each point of these lines corresponds to a different AGN-to-galaxy 
luminosity ratio (that increases from left to right).

In Figure~\ref{oiii_ha}b we do not include the sources classified
as starburst or HII-region galaxies on the basis of the diagnostic 
diagrams because the H$\alpha$ line is likely to be produced within 
the host galaxy rather than by the AGN. We exclude also the sources
with a narrow H$\alpha$ emission line to avoid sources whose H$\alpha$ line 
is contaminated by the emission from the host-galaxy. 
The solid line nicely separates the elusive objects 
affected by large absorption ($>$4$\times$10$^{21}$ cm$^{-2}$) from those
with low absorption ($<$4$\times$10$^{21}$ cm$^{-2}$). More importantly, 
this separating line is fairly independent from the host-galaxy type even
when the host galaxy light dominates the total spectrum 
(unlike the $\Delta$/[OIII]$\lambda$5007\AA\ plot). 
Therefore, Figure~\ref{oiii_ha}  can be used as diagnostic tool to separate between
type~1 and type~2 AGN, as defined in the XBS sample, when 
dilution from the host-galaxy does not allow to apply the usual diagnostic
criteria and when X-ray data are not available.

\section{The catalog}
The result of the spectral classification of the XBS sources is
summarized in Table~\ref{breakdown} while in
Table~\ref{table} we report the relevant optical information for each 
object.  We note that the
classification of the XBS sources has been presented in part in Della Ceca et 
al. (2004). The classification presented in that paper  has been 
revised and refined to take into account the complexity of some spectra 
(like the presence of a significant star-light contribution) 
and, therefore, some of the published classifications (20 in total) have
now changed.  Most (14 out of 20) of the sources with a classification different from
that presented in Della Ceca et al. (2004) are optically elusive AGN or ``normal galaxies'' 
and, therefore, the new classification is based on the X-ray spectrum.
In Table~\ref{table} we have flagged the sources for which
the classification presented here differs from that published in Della Ceca
et al. (2004). 

In Table~\ref{table} we have also listed an optical magnitude for each optical
counterpart. As already discussed, we have not carried out a systematic
photometric follow-up of the XBS sources and, therefore, the magnitudes 
are not homogeneous being taken from different catalogues/observations. 
For about half of the objects (172 objects) we have collected a red (R or r) 
magnitude either from our own observations or from existing catalogues (mostly
the SDSS catalogue). Some of the R magnitudes derived from our own observations 
have been computed from the optical spectrum. 
Another substantial fraction of magnitudes (150) are taken
from the APM facility (we use the red APM filter). For bright (and extended) 
objects the APM magnitude is known to suffer from a large systematic error.
In these cases we have applied the correction described in March\~a et al. 
(2001) to compensate for this systematic error.
Finally, for 26 objects classified as stars we have given the magnitude
V or B present in Simbad.

\section{The classification breakdown}

%                                     One column figure (place early!)
%______________________________________________ Gamma_1 (lg rho, lg e)
   \begin{figure}
   \centering
    \includegraphics[width=9cm]{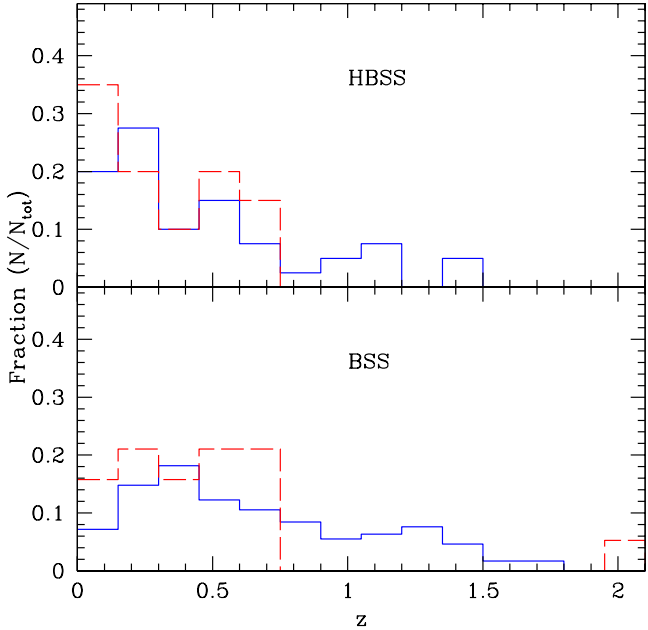}

   \caption{Redshift distribution of the type~1 (solid line) 
and type~2 (dashed line) AGNs in the two
samples (BSS and HBSS).}
              \label{zdist}
    \end{figure}

%

%                                     One column figure (place early!)
%______________________________________________ Gamma_1 (lg rho, lg e)
   \begin{figure}
   \centering
    \includegraphics[width=9cm]{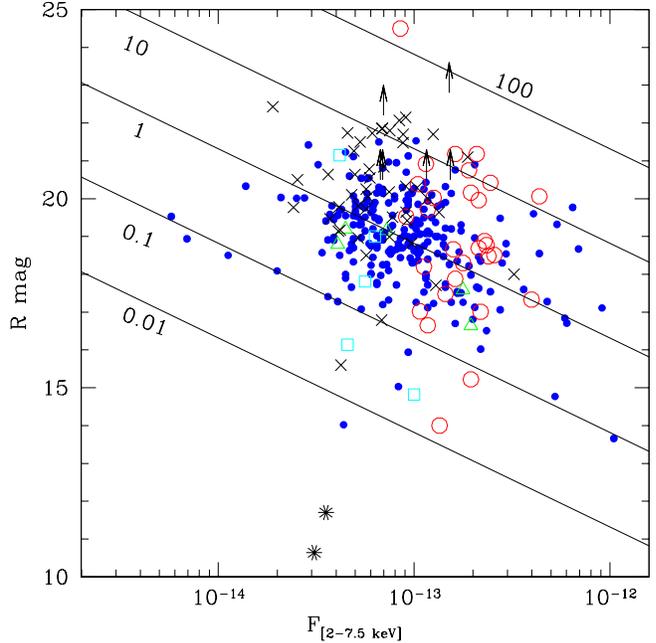}

   \caption{Magnitude vs. 2-7.5 keV flux (as derived from the
count-rates) plot of the XBS extragalactic objects plus the unidentified 
sources: filled points = type~1 AGN, open 
circles = type~2 AGNs, open triangles = BL Lac objects, open squares = 
clusters of galaxies, stars = normal galaxies, crosses = unidentified sources. 
The continuous lines indicate the region of constant X-ray-to-optical flux 
ratio, from 0.01 to 100.}
              \label{x_optical}
    \end{figure}

Table~\ref{breakdown} reports the current classification breakdown of the
sources in the BSS and HBSS samples. Given the high identification
level (87\% and 97\% for the BSS and the HBSS samples respectively) the
numbers in Table~\ref{breakdown} should reflect the true relative compositions 
of the two samples. The first obvious consideration is that the percentage
of stars decreases dramatically from the BSS sample (17\%) to the HBSS 
sample (3\%). Similarly, the relative fraction of type~2/type~1 AGN 
is significantly different in the 2 samples, being a factor 6 higher in the
HBSS (0.48) than  in the BSS (0.08). As expected, the 4.5-7.5 keV energy
band is much more efficient in selecting type~2 AGN (efficiency $\sim$29\%) 
when compared to the softer 0.5-4.5 keV band (efficiency$\sim$6\%). 
It must be noted, however, that the optical recognition of the AGN in the hard
energy band is more difficult when compared to the 0.5-4.5 keV band, 
since about 21\% of the AGN are elusive 
(while only 10\% of the AGN in the BSS are elusive).
The different impact of the problem of dilution on
type~1 and type~2 AGN and on different selection bands should be kept
in mind when deriving statistical considerations on the populations of
AGNs present in X-ray surveys.

As far as the BL Lac objects are concerned the selection efficiency 
in the 0.5-4.5 keV band is about 1-2\%. If this efficiency was the
same in the 4.5-7.5 keV band we would expect $\sim$1 BL Lac,  
something that is statistically consistent with the fact that
no BL Lacs are observed in the HBSS sample.

The redshift distribution of type~1 and type~2 AGN in the two samples 
is shown in Figure~\ref{zdist}. In the BSS sample, the mean redshift of 
type~1 AGNs 
($<z>_{Ty1}$=0.69$\pm$0.03) is significantly different from the 
mean redshift of type~2 AGNs ($<z>_{Ty2}$=0.47$\pm$0.10) while they
are closer  in the HBSS sample ($<z>_{Ty1}$=0.47$\pm$0.06 
$<z>_{Ty2}$=0.33$\pm$0.05).
A K-S test confirms that the z-distribution
of the two classes of AGN are consistent with being derived from the
same parent distribution when considering the HBSS sample (K-S 
probability=33\%)
while they are significantly different (at 95\% confidence level) when 
considering the BSS sample (K-S probability=1.6\%). 
This result probably reflects
the fact that the hard-energy (4.5-7.5 keV) selection is less biased in 
respect to the obscuration (at least in the Compton-thin regime) when
compared to a softer (0.5-4.5 keV) energy selection.

Finally, in Fig.~\ref{x_optical} we plot the extragalactic XBS sources and
the unidentified objects on the magnitude/X-ray flux diagram. 
The identified extragalactic 
sources, with the exception of three objects, have an X-ray-to-optical 
flux ratio (X/O) between 0.005 and 20. At the two  ``extreme'' sides of
the distribution we find the two ``normal'' galaxies, 
that have the lowest values of 
X/O ($\sim$10$^{-4}$) similar to those observed in some 
stars, and, on the other side of the distribution, the high z type~2 QSO, 
discussed in Severgnini et al. (2006), which has the highest value of
X/O ($\sim$200). Among the unidentified sources we
have at least one object whose lower limit on the magnitude (R$>$22.8) 
implies a X/O greater than 60, making it
an excellent candidate of high-z   type~2 QSO. 

Interestingly enough, among the high ($>$10) X/O sources 
3 type~1 AGN are found. These objects  represent a non negligible fraction 
considering that about half of the high X/O sources are still unidentified
and more cases like these may show up after the completion of the spectroscopic
follow-up. A significant presence of type~1 AGN among high X/O sources 
has been found also at lower X-ray fluxes ($\sim$10$^{-14}$ erg s$^{-1}$ 
cm$^{-2}$) in the {\it XMM-Newton} Medium sensitivity Survey (XMS, 
Barcons et al. 2007).

%______________________________________________________________
%
%_____________________________________________________________
%                                             Simple A&A Table
%_____________________________________________________________
%
\begin{table}
\caption{Breakdown of the optical classification. 
Numbers between parenthesis 
indicate the number of sources for which the 
classification  is based on the X-ray spectral analysis 
(see Caccianiga et al. 2007) 
}% title of Table
\label{breakdown}      % is used to refer this table in the text
\centering                          % used for centering table
\begin{tabular}{c c c c }        % centered columns (4 columns)
\hline\hline                 % inserts double horizontal lines
Type & Number & in BSS & in HBSS \\    % table heading 
\hline                        % inserts single horizontal line

AGN 1 & 245 (20) & 244 (20) & 42 (4)\\
AGN 2 &  29 (12) & 19 (5) & 20 (9)\\
AGN (uncertain type) & 1 (1) & 1 (1) & 0 \\ 
BL Lacs & 5 & 5 & 0 \\
``normal'' Galaxies & 2 (2) & 2 (2) & 0\\ 
Clusters of galaxies & 8 & 8 & 1 \\
stars & 58 & 58 & 2 \\
\hline       
IDs       & 348 (35) & 337 (28) & 65 (13)\\
total     & 400     & 389 & 67    \\ 
\hline                            %inserts single line
\end{tabular}

\end{table}

\section{Summary and conclusions}
We have presented the details of the identification work of the sources
in the XBS survey, which is composed by two complete flux limited
samples, the BSS and the HBSS sample, selected in the 0.5-4.5 keV and
4.5-7.5 keV band respectively.
We have secured a redshift and a spectroscopic classification for 348 
(including data from the literature) out of 400 sources,  corresponding to 
87\% of the total list of sources and to 87\% and 97\% considering
the BSS and HBSS samples separately. 

The  results of the identification work can be summarized as follows:

\begin{itemize}
 
\item 
We have quantified the criteria used to
distinguish optically absorbed AGN (i.e. type~2) from optically
non-absorbed (or moderately absorbed) AGN (type~1) and we have shown that 
the adopted dividing line between the two classes of sources 
corresponds to an optical extinction of A$_V\sim$2 mag, which
translates into an expected column density of N$_H\sim$4$\times$10$^{21}$ 
cm$^{-2}$, assuming a Galactic A$_V$/N$_H$ relationship. 

\item
About 10\% of the extragalactic sources (35 objects in total) show an optical 
spectrum which is highly contaminated by the star-light from the host galaxy.
These sources have been studied in detail in a companion paper 
(Caccianiga et al. 2007). Using the X-ray data we have found an elusive
AGN in 33 of 
these objects and we have classified them into type~1 and type~2 AGN 
according to the value of N$_H$ measured from the X-ray spectrum. 
To this end, we have used a N$_H$=4$\times$10$^{21}$ cm$^{-2}$ dividing 
value  which matches (assuming the standard Galactic A$_V$/N$_H$ relation)
the value of A$_V$ (=2 mag) adopted with the optical classification.

\item
We have then proposed two simple diagnostic diagrams. The first one, 
based on the 4000\AA\ break and the [OIII]$\lambda$5007\AA\ equivalent width, can reliably 
distinguish between type~1 and type~2 AGN if the host galaxy does not 
dominate the optical spectrum. The second uses the H$\alpha$ and 
[OIII]$\lambda$5007\AA\ line equivalent widths to classify into type~1 
and type~2 the elusive AGN sources in which a possibly broad H$\alpha$ emission
line is detected.

\item
We find that the AGN represent the most numerous population  at the 
flux limit of the XBS survey ($\sim$10$^{-13}$ erg cm$^{-2}$ s$^{-1}$) 
constituting 80\% of the  XBS sources selected in the 0.5-4.5 keV energy 
band and 95\% of the ``hard'' (4.5-7.5 keV) selected objects. Galactic 
sources populate significantly the 0.5-4.5 keV sample (17\%) and only 
marginally (3\%) the 4.5-7.5 keV sample. The remaining sources in both samples 
are clusters/groups of galaxies and normal galaxies (i.e. probably 
not powered by an AGN). 

\item
As expected, the percentage of type~2 AGN dramatically increases going from the 0.5-4.5 keV sample 
(f=N$_{AGN 2}$/N$_{AGN}$=7\%) to the 4.5-7.5 keV sample (f=32\%). A detailed analysis
on the intrinsic (i.e. taking into account the selection effects) relative fraction of 
type~1 and type~2 AGN will be be presented in a forthcoming paper (Della Ceca et al. 2007, in prep.).

\end{itemize}

\begin{acknowledgements}
We thank the referee for useful suggestions. 
Based on observations made with: ESO Telescopes at the La Silla  and Paranal 
Observatories 
under programme IDs: 069.B-0035, 070.A-0216, 074.A-0024, 075.B-0229, 
076.A-0267;  
the Italian Telescopio Nazionale Galileo (TNG) operated on the island of 
La Palma by 
the Fundaci\'on Galileo Galilei of the INAF (Istituto Nazionale di Astrofisica)
 at the 
Spanish Observatorio del Roque de los Muchachos of the Instituto de Astrofisica
de Canarias; the German-Spanish
Astronomical Center, Calar Alto (operated jointly by Max-Planck
Institut  f\"{u}r Astronomie and Instututo de Astrofisica de
Andalucia, CSIC). AC, RDC, TM and PS acknowledge financial support from the 
MIUR, grant PRIN-MUR 2006-02-5203 and from the Italian Space Agency (ASI),
grants n. I/088/06/0 and n. I/023/05/0. 
This research has made use of the Simbad database and of the 
NASA/IPAC Extragalactic Database
(NED) which is operated by the Jet Propulsion Laboratory,
California Institute of Technology, under contract with the
National Aeronautics and Space Administration. 
The research described in this 
paper has been conducted within  the {\it XMM-Newton Survey Science Center} 
(SSC, see http://xmmssc-www.star.le.ac.uk.) collaboration, involving a 
consortium of 
10 institutions, appointed 
by ESA to help the SOC in developing the software analysis system, 
to pipeline process all the {\it XMM-Newton} data, and to exploit the 
{\it XMM-Newton} serendipitous detections. 
\end{acknowledgements}

\newpage

\begin{table*}
\caption{Optical properties of the XBSS sources}
\label{table}
\begin{tabular}{l c l l l l r r c}
\hline\hline
name & Sample & Optical position & Class & flag class & z & mag & flag mag & reference \\ 
     &        &     (J2000)      &       &      &   &     &  &  \\
\hline
XBSJ000027.7--250442 & bss & 00 00 27.68 --25 04 42.8  & AGN1 &        & 0.336 & 19.0 & 3 & 1          \\
XBSJ000031.7--245502 & bss & 00 00 31.89 --24 54 59.5  & AGN1 &        & 0.284 & 17.2 & 1 & 1          \\
XBSJ000100.2--250501 & bss & 00 01 00.23 --25 05 01.5  & AGN1 &        & 0.850 & 20.4 & 3 & 1          \\
XBSJ000102.4--245850 & bss & 00 01 02.46 --24 58 49.6  & AGN1 &        & 0.433 & 20.3 & 1 & 1          \\
XBSJ000532.7+200716 & bss & 00 05 32.84 +20 07 17.4  & AGN1 & 1      3 & 0.119 & 17.9 & 3 & obs        \\
XBSJ001002.4+110831 & bss & 00 10 02.66 +11 08 34.4  & star &        & -- &  5.5 & 5 & 43         \\
XBSJ001051.6+105140 & bss & 00 10 51.41 +10 51 40.5  & star &        & -- & 15.8 & 4 & obs        \\
XBSJ001749.7+161952 & bss & 00 17 49.93 +16 19 56.1  & star &        & -- &  7.2 & 5 & 43         \\
XBSJ001831.6+162925 & bss & 00 18 32.02 +16 29 25.9  & AGN1 &        & 0.553 & 18.3 & 3 & 42,2       \\
XBSJ002618.5+105019 & bss,hbss & 00 26 18.71 +10 50 19.6  & AGN1 &        & 0.473 & 17.5 & 3 & obs        \\
XBSJ002637.4+165953 & bss & 00 26 37.46 +16 59 54.4  & AGN1 &        & 0.554 & 18.9 & 3 & obs        \\
XBSJ002707.5+170748 & bss & 00 27 07.78 +17 07 50.5  & AGN1 &        & 0.930 & 20.2 & 1 & obs        \\
XBSJ002953.1+044524 & bss & 00 29 53.16 +04 45 24.1  & star &        & -- &  9.5 & 6 & 43         \\
XBSJ003255.9+394619 & bss & 00 32 55.73 +39 46 19.4  & AGN1 &        & 1.139 & 17.7 & 3 & obs        \\
XBSJ003315.5--120700 & bss & 00 33 15.63 --12 06 58.7  & AGN1 &        & 1.206 & 19.8 & 3 & obs        \\
XBSJ003316.0--120456 & bss & 00 33 16.04 --12 04 56.2  & AGN1 &        & 0.660 & 18.9 & 3 & obs        \\
XBSJ003418.9--115940 & bss & 00 34 19.00 --11 59 38.2  & AGN1 &        & 0.850 & 20.6 & 1 & obs        \\
XBSJ005009.9--515934 & bss & 00 50 09.66 --51 59 32.4  & AGN1 &        & 0.610 & 20.1 & 3 & obs        \\
XBSJ005031.1--520012 & bss & 00 50 30.85 --52 00 09.8  & AGN1 &        & 0.463 & 18.7 & 3 & obs        \\
XBSJ005032.3--521543 & bss & 00 50 32.13 --52 15 42.3  & AGN1 &        & 1.216 & 19.9 & 3 & obs        \\
XBSJ005822.9--274016 & bss & 00 58 22.96 --27 40 14.2  & star &        & -- & 12.3 & 5 & 43         \\
XBSJ010421.4--061418 & bss & 01 04 21.57 --06 14 17.5  & AGN1 &        & 0.520 & 21.2 & 2 & obs        \\
XBSJ010432.8--583712 & bss & 01 04 32.64 --58 37 11.2  & AGN1 &        & 1.640 & 19.3 & 3 & obs        \\
XBSJ010701.5--172748 & bss & 01 07 01.47 --17 27 46.4  & AGN1 &        & 0.890 & 19.2 & 3 & obs        \\
XBSJ010747.2--172044 & bss & 01 07 47.50 --17 20 42.0  & AGN1 &        & 0.980 & 17.5 & 3 & obs        \\
XBSJ012000.0--110429 & bss & 01 20 00.10 --11 04 30.0  & AGN1 &        & 0.351 & 20.3 & 3 & obs        \\
XBSJ012025.2--105441 & bss & 01 20 25.31 --10 54 38.6  & AGN1 &        & 1.338 & 18.9 & 3 & 3,39       \\
XBSJ012057.4--110444 & bss & 01 20 57.38 --11 04 44.0  & AGN2 &        & 0.072 & 16.7 & 1 & obs        \\
XBSJ012119.9--110418 & bss & 01 21 19.99 --11 04 14.9  & AGN1 &        & 0.204 & 17.5 & 4 & obs        \\
XBSJ012505.4+014624 & bss & 01 25 05.50 +01 46 27.2  & AGN1 &        & 1.567 & 19.0 & 3 & obs        \\
XBSJ012540.2+015752 & bss & 01 25 40.36 +01 57 53.8  & AGN1 & 1      3 & 0.123 & 17.3 & 4 & obs        \\
XBSJ012654.3+191246 & bss & 01 26 54.45 +19 12 52.5  & AGN1 & 1      3 & 0.043 & 13.7 & 1 & obs        \\
XBSJ012757.2+190000 & bss & 01 27 57.05 +19 00 02.0  & star &        & -- & 12.7 & 5 & 41         \\
XBSJ012757.3+185923 & bss & 01 27 57.24 +18 59 26.3  & star &        & -- &  9.4 & 5 & 43         \\
XBSJ013204.9--400050 & bss & 01 32 05.19 --40 00 48.2  & AGN1 &        & 0.450 & 19.1 & 3 & obs        \\
XBSJ013240.1--133307 & bss,hbss & 01 32 40.29 --13 33 06.5  & AGN2 &        & 0.562 & 20.0 & 3 & obs        \\
XBSJ013811.7--175416 & bss & 01 38 11.72 --17 54 13.4  & BL &        & 0.530 & 19.2 & 3 & obs        \\
XBSJ013944.0--674909 & bss,hbss & 01 39 43.70 --67 49 08.1  & AGN1 & 1       & 0.104 & 17.7 & 4 & obs        \\
XBSJ014100.6--675328 & bss,hbss & 01 41 00.29 --67 53 27.5  & star &        & -- & 16.4 & 3 & obs        \\
XBSJ014109.9--675639 & bss & 01 41 09.53 --67 56 38.7  & AGN1 & 1       & 0.226 & 19.2 & 3 & obs        \\
XBSJ014227.0+133453 & bss & 01 42 27.31 +13 34 53.1  & AGN1 & 1      3 & 0.275 & 19.3 & 2 & obs        \\
XBSJ014251.5+133352 & bss & 01 42 51.72 +13 33 52.7  & AGN1 &        & 1.071 & 19.0 & 2 & obs        \\
XBSJ015916.9+003010 & bss & 01 59 17.20 +00 30 13.0  & CL &        & 0.382 & 17.8 & 2 & obs        \\
XBSJ015957.5+003309 & bss,hbss & 01 59 57.65 +00 33 10.8  & AGN1 &        & 0.310 & 18.8 & 2 & obs        \\
XBSJ020029.0+002846 & bss & 02 00 29.07 +00 28 46.7  & AGN1 &        & 0.174 & 18.0 & 2 & obs        \\
XBSJ020757.3+351828 & bss & 02 07 57.15 +35 18 28.2  & AGN1 &        & 0.188 & 18.3 & 3 & obs        \\
XBSJ020845.1+351438 & bss & 02 08 44.96 +35 14 37.2  & AGN1 &        & 0.415 & 19.1 & 3 & obs        \\
XBSJ021640.7--044404 & bss,hbss & 02 16 40.72 --04 44 04.8  & AGN1 &        & 0.873 & 17.2 & 4 & obs        \\
XBSJ021642.3--043553 & bss & 02 16 42.36 --04 35 51.9  & AGN2 &        & 1.985 & 24.5 & 1 & obs        \\
XBSJ021808.3--045845 & bss,hbss & 02 18 08.24 --04 58 45.2  & AGN1 &        & 0.712 & 17.7 & 3 & 40         \\
XBSJ021817.4--045113 & bss,hbss & 02 18 17.45 --04 51 12.4  & AGN1 &        & 1.080 & 19.5 & 3 & 40         \\
XBSJ021820.6--050427 & bss & 02 18 20.46 --05 04 26.2  & AGN1 &        & 0.646 & 18.7 & 3 & obs        \\
XBSJ021822.2--050615 & hbss & 02 18 22.16 --05 06 14.4  & AGN2 & 1       & 0.044 & 15.2 & 1 & obs        \\
XBSJ021830.0--045514 & bss & 02 18 29.91 --04 55 13.8  & star &        & -- & 14.3 & 1 & 40         \\
XBSJ021923.2--045148 & bss & 02 19 23.30 --04 51 48.6  & AGN1 &        & 0.632 & 18.9 & 3 & obs        \\
XBSJ022253.0--044515 & bss & 02 22 53.15 --04 45 13.1  & AGN1 &        & 1.420 & 20.5 & 1 & obs        \\
\hline
\end{tabular}
\end{table*}
\newpage
\addtocounter{table}{-1}
\begin{table*}
\caption{continue}
\begin{tabular}{l c l l l l r r c}
name & Sample & Optical position & Class & flag class & z & mag & flag mag & reference \\ 
     &        &     (J2000)      &       &      &   &     &  &  \\
\hline
XBSJ022707.7--050819 & bss & 02 27 07.93 --05 08 17.4  & AGN2 &        & 0.358 & 18.9 & 3 & obs        \\
XBSJ023459.7--294436 & bss & 02 34 59.97 --29 44 34.6  & AGN1 &        & 0.446 & 17.7 & 3 & obs        \\
XBSJ023530.2--523045 & bss & 02 35 30.38 --52 30 43.2  & AGN1 &        & 0.429 & 18.8 & 3 & obs        \\
XBSJ023713.5--522734 & bss,hbss & 02 37 13.57 --52 27 34.1  & AGN1 &        & 0.193 & 17.1 & 3 & obs        \\
XBSJ023853.2--521911 & bss & 02 38 53.41 --52 19 09.9  & AGN1 &        & 0.648 & 19.4 & 3 & obs        \\
XBSJ024200.9+000020 & bss & 02 42 00.91 +00 00 21.1  & AGN1 &    2    & 1.112 & 18.4 & 2 & 4          \\
XBSJ024204.7+000814 & bss & 02 42 04.77 +00 08 14.7  & AGN1 &        & 0.383 & 18.9 & 2 & obs        \\
XBSJ024325.6--000413 & bss & 02 43 25.50 --00 04 13.0  & AGN1 &        & 0.356 & 19.4 & 3 & obs        \\
XBSJ025606.1+001635 & bss & 02 56 06.00 +00 16 34.8  & AGN1 &        & 0.629 & 20.1 & 2 & obs        \\
XBSJ025645.4+000031 & bss & 02 56 45.29 +00 00 33.2  & AGN1 & 1       & 0.359 & 19.3 & 2 & obs        \\
XBSJ030206.8--000121 & bss,hbss & 03 02 06.77 --00 01 21.1  & AGN1 &    2    & 0.641 & 18.8 & 3 & 6          \\
XBSJ030614.1--284019 & bss,hbss & 03 06 14.17 --28 40 20.1  & AGN1 &        & 0.278 & 18.5 & 3 & obs        \\
XBSJ030641.0--283559 & bss & 03 06 41.10 --28 35 58.8  & AGN1 &        & 0.367 & 17.3 & 3 & obs        \\
XBSJ031015.5--765131 & bss,hbss & 03 10 15.69 --76 51 32.9  & AGN1 &        & 1.187 & 17.6 & 3 & 7          \\
XBSJ031146.1--550702 & bss,hbss & 03 11 46.08 --55 07 00.2  & AGN2 &        & 0.162 & 17.3 & 3 & obs        \\
XBSJ031311.7--765428 & bss & 03 13 11.85 --76 54 30.4  & AGN1 &        & 1.274 & 19.1 & 3 & 7          \\
XBSJ031401.3--545959 & bss & 03 14 01.37 --54 59 56.4  & AGN1 &        & 0.841 & 20.2 & 3 & 8          \\
XBSJ031549.4--551811 & bss & 03 15 49.60 --55 18 13.0  & AGN1 &        & 0.808 & 20.3 & 3 & 8          \\
XBSJ031851.9--441815 & bss & 03 18 52.04 --44 18 16.7  & AGN1 &        & 1.360 & 19.0 & 3 & obs        \\
XBSJ031859.2--441627 & bss,hbss & 03 18 59.46 --44 16 26.4  & AGN1 & 1       & 0.140 & 16.7 & 1 & obs        \\
XBSJ033208.7--274735 & bss & 03 32 08.67 --27 47 34.3  & AGN1 &       3 & 0.544 & 18.3 & 3 & 9,10       \\
XBSJ033226.9--274107 & bss & 03 32 27.03 --27 41 04.8  & AGN1 &        & 0.736 & 18.8 & 3 & obs        \\
XBSJ033435.5--254259 & bss & 03 34 35.76 --25 42 54.9  & AGN1 &        & 1.190 & 19.4 & 3 & obs        \\
XBSJ033453.9--254154 & bss & 03 34 54.14 --25 41 53.2  & AGN1 &        & 1.160 & 18.6 & 3 & obs        \\
XBSJ033506.0--255619 & bss & 03 35 06.02 --25 56 19.3  & AGN1 &        & 1.430 & 17.4 & 3 & obs        \\
XBSJ033845.7--352253 & hbss & 03 38 46.01 --35 22 52.2  & AGN2 &        & 0.113 & 17.0 & 3 & obs        \\
XBSJ033851.4--352646 & bss & 03 38 51.60 --35 26 44.7  & AGN1 &        & 1.070 & 19.5 & 1 & obs        \\
XBSJ033912.1--352813 & bss & 03 39 12.18 --35 28 12.4  & AGN1 &        & 0.466 & 19.7 & 1 & obs        \\
XBSJ033942.8--352411 & bss & 03 39 42.90 --35 24 10.3  & AGN1 &        & 1.043 & 19.0 & 3 & 11         \\
XBSJ040658.8--712457 & hbss & 04 06 58.85 --71 24 59.6  & AGN2 &        & 0.181 & 18.7 & 3 & obs        \\
XBSJ040744.6--710846 & bss & 04 07 44.56 --71 08 47.5  & star &        & -- & 17.9 & 3 & obs        \\
XBSJ040758.9--712833 & hbss & 04 07 58.53 --71 28 32.9  & AGN2 &        & 0.134 & 17.0 & 3 & obs        \\
XBSJ040807.2--712702 & bss & 04 08 07.08 --71 27 01.6  & star &        & -- & 12.4 & 4 & 43         \\
XBSJ041108.1--711341 & bss,hbss & 04 11 08.59 --71 13 43.0  & AGN1 &        & 0.923 & 20.3 & 1 & obs        \\
XBSJ043448.3--775329 & bss & 04 34 47.78 --77 53 28.3  & AGN1 & 1       & 0.097 & 17.7 & 3 & obs        \\
XBSJ045942.4+015843 & bss & 04 59 42.50 +01 58 44.2  & AGN1 &        & 0.248 & 19.1 & 3 & obs        \\
XBSJ050011.7+013948 & bss & 05 00 11.72 +01 39 48.8  & AGN1 &        & 0.360 & 19.9 & 3 & obs        \\
XBSJ050446.3--283821 & bss & 05 04 46.38 --28 38 20.1  & AGN1 &        & 0.840 & 20.6 & 1 & obs        \\
XBSJ050453.4--284532 & bss & 05 04 53.35 --28 45 31.0  & AGN1 & 1       & 0.204 & 19.0 & 1 & obs        \\
XBSJ050501.8--284149 & bss & 05 05 01.90 --28 41 48.5  & AGN1 &        & 0.257 & 18.8 & 1 & obs        \\
XBSJ050536.6--290050 & bss,hbss & 05 05 36.56 --29 00 49.7  & AGN2 &        & 0.577 & 21.2 & 1 & obs        \\
XBSJ051617.1+794408 & bss & 05 16 17.23 +79 44 11.0  & star &        & -- &  9.3 & 5 & 43         \\
XBSJ051655.3--104104 & bss & 05 16 55.28 --10 41 02.4  & AGN1 &        & 0.568 & 20.3 & 1 & obs        \\
XBSJ051822.6+793208 & bss & 05 18 22.55 +79 32 09.8  & AGN1 & 1      3 & 0.053 & 15.0 & 4 & obs        \\
XBSJ051955.5--455727 & bss & 05 19 55.56 --45 57 25.2  & AGN1 &        & 0.562 & 19.0 & 1 & obs        \\
XBSJ052022.0--252309 & bss & 05 20 22.17 --25 23 10.5  & AGN1 &        & 0.745 & 19.8 & 1 & obs        \\
XBSJ052048.9--454128 & bss & 05 20 49.30 --45 41 30.2  & star &        & -- & 11.9 & 5 & 43         \\
XBSJ052108.5--251913 & bss,hbss & 05 21 08.71 --25 19 13.3  & AGN1 &        & 1.196 & 17.5 & 3 & obs        \\
XBSJ052116.2--252957 & bss & 05 21 16.08 --25 29 58.3  & AGN1 & 1       & 0.332 & 19.6 & 1 & obs        \\
XBSJ052128.9--253032 & hbss & 05 21 29.04 --25 30 32.3  & AGN2 & 1       & 0.588 & 20.8 & 1 & obs        \\
XBSJ052144.1--251518 & bss & 05 21 44.37 --25 15 23.0  & AGN1 &        & 0.321 & 18.9 & 1 & obs        \\
XBSJ052155.0--252200 & bss & 05 21 55.32 --25 22 00.9  & star &        & -- & 13.0 & 1 & 41         \\
XBSJ052509.3--333051 & bss & 05 25 09.29 --33 30 52.9  & CL &        & 0.704 & 21.2 & 1 & obs        \\
XBSJ052543.6--334856 & bss & 05 25 43.61 --33 48 57.5  & AGN1 &        & 0.735 & 19.7 & 1 & obs        \\
XBSJ061342.7+710725 & bss & 06 13 43.20 +71 07 24.6  & BL &        & 0.267 & 16.6 & 4 & 12         \\
XBSJ062134.8--643150 & bss & 06 21 34.74 --64 31 51.5  & AGN1 &        & 1.277 & 18.0 & 3 & obs        \\
\hline
\end{tabular}
\end{table*}
\newpage
\addtocounter{table}{-1}
\begin{table*}
\caption{continue}
\begin{tabular}{l c l l l l r r c}
name & Sample & Optical position & Class & flag class & z & mag & flag mag & reference \\ 
     &        &     (J2000)      &       &      &   &     &  &  \\
\hline
XBSJ062425.7--642958 & bss & 06 24 25.78 --64 29 58.3  & star &        & -- & 11.1 & 5 & 43         \\
XBSJ065214.1+743230 & bss & 06 52 14.62 +74 32 29.4  & AGN1 &        & 0.620 & 19.9 & 3 & obs        \\
XBSJ065237.4+742421 & bss & 06 52 37.62 +74 24 20.2  & CL &        & 0.360 & 19.0 & 1 & obs        \\
XBSJ065400.0+742045 & bss & 06 54 00.28 +74 20 44.0  & AGN1 &        & 0.362 & 19.3 & 1 & obs        \\
XBSJ065744.3--560817 & bss & 06 57 44.17 --56 08 18.8  & AGN1 &        & 0.120 & 17.1 & 1 & obs        \\
XBSJ065839.5--560813 & bss & 06 58 39.33 --56 08 12.2  & AGN1 &        & 0.211 & 17.3 & 1 & obs        \\
XBSJ074202.7+742625 & bss,hbss & 07 42 02.68 +74 26 24.7  & AGN1 &        & 0.599 & 20.9 & 1 & obs        \\
XBSJ074312.1+742937 & bss,hbss & 07 43 12.60 +74 29 36.3  & AGN1 &        & 0.312 & 17.1 & 4 & 13         \\
XBSJ074338.7+495431 & bss & 07 43 38.99 +49 54 28.5  & AGN1 &        & 0.221 & 19.2 & 2 & obs        \\
XBSJ074352.0+744258 & bss & 07 43 52.98 +74 42 57.9  & AGN1 &        & 0.800 & 18.8 & 2 & 40         \\
XBSJ074359.7+744057 & bss & 07 44 00.55 +74 40 56.5  & star &        & -- & 14.6 & 4 & obs        \\
XBSJ075117.9+180856 & bss & 07 51 17.96 +18 08 56.0  & AGN1 & 1       & 0.255 & 18.7 & 2 & obs        \\
XBSJ080309.8+650807 & bss & 08 03 09.11 +65 08 06.7  & star &        & -- &  7.7 & 5 & 43         \\
XBSJ080608.1+244420 & bss & 08 06 08.15 +24 44 21.3  & AGN1 &        & 0.357 & 18.3 & 2 & obs        \\
XBSJ083737.0+255151 & bss,hbss & 08 37 37.04 +25 51 51.6  & AGN1 & 1      3 & 0.105 & 16.5 & 2 & obs        \\
XBSJ083737.1+254751 & bss,hbss & 08 37 37.08 +25 47 50.5  & AGN1 &        & 0.080 & 16.8 & 2 & 13,39      \\
XBSJ083838.6+253616 & bss & 08 38 38.48 +25 36 17.1  & AGN1 &        & 0.601 & 19.2 & 2 & obs        \\
XBSJ083905.9+255010 & bss & 08 39 05.91 +25 50 09.3  & AGN1 &        & 0.250 & 20.0 & 2 & obs        \\
XBSJ084026.2+650638 & bss & 08 40 26.11 +65 06 38.3  & AGN1 &        & 1.144 & 18.7 & 1 & obs        \\
XBSJ084651.7+344634 & bss & 08 46 51.68 +34 46 34.7  & AGN1 &        & 1.115 & 18.1 & 3 & 42         \\
XBSJ085427.8+584158 & bss & 08 54 28.24 +58 42 05.3  & star &        & -- & 10.0 & 5 & 43         \\
XBSJ085530.7+585129 & bss & 08 55 30.97 +58 51 29.0  & AGN1 &        & 0.905 & 21.4 & 2 & obs        \\
XBSJ090729.1+620824 & bss & 09 07 29.30 +62 08 27.0  & AGN2 & 1      3 & 0.388 & 20.4 & 2 & obs        \\
XBSJ091043.4+054757 & bss & 09 10 43.33 +05 48 01.8  & star &        & -- & 17.5 & 2 & obs        \\
XBSJ091828.4+513931 & bss,hbss & 09 18 28.59 +51 39 32.3  & AGN1 &        & 0.185 & 17.1 & 2 & obs        \\
XBSJ094526.2--085006 & bss & 09 45 26.25 --08 50 05.9  & AGN1 & 1       & 0.314 & 18.2 & 1 & obs        \\
XBSJ094548.3--084824 & bss & 09 45 48.18 --08 48 23.7  & AGN1 &        & 1.748 & 18.6 & 3 & obs        \\
XBSJ095054.5+393924 & bss & 09 50 54.88 +39 39 27.4  & AGN1 &        & 1.299 & 19.6 & 2 & obs        \\
XBSJ095218.9--013643 & bss,hbss & 09 52 19.08 --01 36 43.4  & AGN1 &        & 0.020 & 13.7 & 4 & 14         \\
XBSJ095309.7+013558 & bss & 09 53 10.13 +01 35 56.6  & AGN1 &        & 0.477 & 19.3 & 1 & obs        \\
XBSJ095341.1+014204 & bss & 09 53 41.36 +01 42 02.4  & CL &        & 0.090 & 14.8 & 2 & obs        \\
XBSJ095416.9+173627 & bss & 09 54 16.74 +17 36 28.4  & CL &        & -- & 20.2 & 2 & obs        \\
XBSJ095509.6+174124 & bss & 09 55 09.63 +17 41 24.9  & AGN1 &        & 1.290 & 20.1 & 2 & obs        \\
XBSJ095955.2+251549 & bss & 09 59 55.07 +25 15 51.7  & star &        & -- & 11.8 & 2 & obs        \\
XBSJ100032.5+553626 & bss & 10 00 32.29 +55 36 30.6  & AGN2 & 1       & 0.216 & 17.9 & 2 & 15,16,39   \\
XBSJ100100.0+252103 & bss & 10 01 00.12 +25 21 04.9  & AGN1 &        & 0.794 & 19.4 & 2 & obs        \\
XBSJ100309.4+554135 & bss & 10 03 09.45 +55 41 34.5  & AGN1 &        & 0.673 & 19.0 & 2 & 15,16,39   \\
XBSJ100828.8+535408 & bss & 10 08 28.95 +53 54 05.8  & AGN1 &        & 0.384 & 18.7 & 1 & obs        \\
XBSJ100921.7+534926 & bss & 10 09 21.88 +53 49 25.5  & AGN1 &        & 0.387 & 18.9 & 2 & 15,16,39   \\
XBSJ100926.5+533426 & bss & 10 09 26.75 +53 34 24.3  & AGN1 &        & 1.718 & 19.3 & 2 & 15,16,39   \\
XBSJ101506.0+520157 & bss & 10 15 06.05 +52 01 58.2  & AGN1 &        & 0.610 & 19.6 & 2 & obs        \\
XBSJ101511.8+520708 & bss & 10 15 11.96 +52 07 07.2  & AGN1 &        & 0.888 & 20.5 & 2 & obs        \\
XBSJ101706.5+520245 & bss & 10 17 06.69 +52 02 47.2  & BL &        & 0.377 & 18.8 & 2 & obs        \\
XBSJ101838.0+411635 & bss & 10 18 37.99 +41 16 38.3  & AGN1 &        & 0.577 & 19.8 & 2 & obs        \\
XBSJ101843.0+413515 & bss & 10 18 43.16 +41 35 16.5  & AGN1 & 1      3 & 0.084 & 15.9 & 2 & obs        \\
XBSJ101850.5+411506 & bss,hbss & 10 18 50.53 +41 15 08.3  & AGN1 &        & 0.577 & 18.4 & 2 & obs        \\
XBSJ101922.6+412049 & bss,hbss & 10 19 22.73 +41 20 50.1  & AGN1 &        & 0.239 & 18.5 & 2 & obs        \\
XBSJ102044.1+081424 & bss & 10 20 44.17 +08 14 23.8  & star &        & -- & 15.2 & 4 & obs        \\
XBSJ102412.3+042023 & bss & 10 24 12.33 +04 20 25.8  & AGN1 &        & 1.458 & 19.5 & 2 & obs        \\
XBSJ102417.5+041656 & bss & 10 24 17.46 +04 16 57.8  & AGN1 &        & 1.712 & 20.1 & 2 & obs        \\
XBSJ103154.1+310732 & bss & 10 31 54.12 +31 07 31.3  & AGN1 &        & 0.299 & 18.8 & 2 & 40         \\
XBSJ103745.7+532353 & bss & 10 37 45.51 +53 23 53.0  & AGN1 &        & 2.347 & 19.8 & 2 & obs        \\
XBSJ103932.7+205426 & bss & 10 39 32.68 +20 54 27.6  & AGN1 &        & 0.237 & 18.8 & 2 & obs        \\
XBSJ103935.8+533036 & bss & 10 39 35.75 +53 30 38.6  & AGN1 &        & 0.229 & 18.4 & 2 & obs        \\
XBSJ104026.9+204542 & bss,hbss & 10 40 26.84 +20 45 44.5  & AGN1 &        & 0.465 & 19.8 & 2 & obs        \\
XBSJ104425.0--013521 & bss & 10 44 24.87 --01 35 19.5  & AGN1 &        & 1.571 & 19.0 & 3 & 9          \\
\hline
\end{tabular}
\end{table*}
\newpage
\addtocounter{table}{-1}
\begin{table*}
\caption{continue}
\begin{tabular}{l c l l l l r r c}
name & Sample & Optical position & Class & flag class & z & mag & flag mag & reference \\ 
     &        &     (J2000)      &       &      &   &     &  &  \\
\hline
XBSJ104509.3--012442 & bss & 10 45 09.32 --01 24 42.5  & AGN1 &        & 0.472 & 20.0 & 1 & obs        \\
XBSJ104522.1--012843 & bss,hbss & 10 45 22.09 --01 28 44.5  & AGN1 &    2    & 0.782 & 19.4 & 3 & 9          \\
XBSJ104912.8+330459 & bss,hbss & 10 49 12.58 +33 05 01.3  & AGN1 &        & 0.226 & 18.6 & 2 & obs        \\
XBSJ105131.1+573439 & bss & 10 51 31.25 +57 34 38.8  & star &        & -- & 14.7 & 3 & obs        \\
XBSJ105239.7+572431 & bss & 10 52 39.76 +57 24 30.6  & AGN1 &        & 1.113 & 17.8 & 2 & 17,39      \\
XBSJ105316.9+573551 & bss & 10 53 16.97 +57 35 50.1  & AGN1 &        & 1.204 & 18.8 & 3 & 17,39      \\
XBSJ105335.0+572540 & bss & 10 53 35.10 +57 25 42.3  & AGN1 &        & 0.784 & 21.1 & 2 & 17         \\
XBSJ105339.7+573104 & bss & 10 53 39.80 +57 31 03.9  & AGN1 &        & 0.586 & 19.8 & 2 & 17         \\
XBSJ105624.2--033522 & bss & 10 56 24.00 --03 35 26.6  & AGN1 &        & 0.635 & 19.0 & 1 & obs        \\
XBSJ110320.1+355803 & bss & 11 03 20.05 +35 58 04.2  & star &        & -- &  7.5 & 5 & 43         \\
XBSJ111654.8+180304 & bss & 11 16 54.72 +18 03 05.9  & G & 1      3 & 0.003 & 10.6 & 2 & 18         \\
XBSJ111928.5+130250 & bss & 11 19 28.39 +13 02 51.3  & AGN1 &        & 2.394 & 18.0 & 2 & obs        \\
XBSJ111933.0+212756 & bss & 11 19 33.22 +21 27 57.6  & AGN1 &        & 0.282 & 19.2 & 2 & 15,16      \\
XBSJ111942.1+211516 & bss & 11 19 42.14 +21 15 16.8  & AGN1 &        & 1.288 & 20.0 & 2 & 15,16      \\
XBSJ112022.3+125252 & bss & 11 20 22.37 +12 52 50.6  & AGN1 &        & 0.406 & 18.9 & 3 & obs        \\
XBSJ112026.7+431520 & hbss & 11 20 26.62 +43 15 18.2  & AGN2 & 1       & 0.146 & 17.5 & 2 & obs        \\
XBSJ112046.7+125429 & bss & 11 20 46.75 +12 54 29.5  & AGN1 &        & 0.382 & 19.4 & 2 & obs        \\
XBSJ113106.9+312518 & bss,hbss & 11 31 06.94 +31 25 19.6  & AGN1 &        & 1.482 & 19.4 & 2 & obs        \\
XBSJ113121.8+310252 & bss,hbss & 11 31 21.81 +31 02 54.8  & AGN2 &        & 0.190 & 18.5 & 2 & 40         \\
XBSJ113128.6--195903 & bss & 11 31 28.44 --19 59 03.2  & AGN1 &        & 0.363 & 18.6 & 3 & obs        \\
XBSJ113148.7+311358 & bss,hbss & 11 31 48.66 +31 14 01.3  & AGN2 &        & 0.500 & 20.4 & 2 & 42         \\
XBSJ113837.9--373402 & bss & 11 38 37.74 --37 33 59.9  & AGN1 &        & 0.120 & 18.3 & 3 & obs        \\
XBSJ115846.9+551625 & bss & 11 58 47.01 +55 16 24.3  & AGN1 &        & 0.518 & 19.6 & 2 & 15,16      \\
XBSJ120359.1+443715 & bss & 12 03 59.10 +44 37 14.8  & AGN1 &        & 0.641 & 19.5 & 2 & obs        \\
XBSJ120413.7+443149 & bss & 12 04 13.72 +44 31 47.6  & AGN1 &        & 0.492 & 19.9 & 2 & obs        \\
XBSJ122017.5+752217 & bss & 12 20 17.76 +75 22 15.2  & G & 1       & 0.006 & 11.7 & 2 & 19         \\
XBSJ122350.4+752231 & bss & 12 23 50.97 +75 22 28.6  & AGN1 &        & 0.565 & 18.8 & 3 & 40         \\
XBSJ122655.1+012002 & bss & 12 26 54.98 +01 20 00.9  & star &        & -- & 18.2 & 2 & obs        \\
XBSJ122656.5+013126 & bss,hbss & 12 26 56.46 +01 31 24.4  & AGN2 &        & 0.733 & 20.2 & 2 & obs        \\
XBSJ122658.1+333246 & bss & 12 26 58.20 +33 32 49.0  & CL &        & 0.891 & 20.6 & 2 & 20,21      \\
XBSJ122751.2+333842 & bss & 12 27 51.17 +33 38 46.5  & star &        & -- & 15.4 & 4 & obs        \\
XBSJ122837.3+015720 & bss & 12 28 37.24 +01 57 19.5  & star &        & -- & 13.4 & 2 & 42         \\
XBSJ122942.3+015525 & bss & 12 29 42.48 +01 55 24.9  & star &        & -- & 14.4 & 2 & obs        \\
XBSJ123116.5+641115 & bss & 12 31 16.50 +64 11 14.4  & AGN1 &        & 0.454 & 20.8 & 2 & 40         \\
XBSJ123208.7+640304 & bss & 12 32 08.89 +64 03 02.6  & star &        & -- & 14.8 & 2 & obs        \\
XBSJ123218.5+640311 & bss & 12 32 18.83 +64 03 09.8  & AGN1 &        & 1.013 & 21.0 & 2 & 40         \\
XBSJ123538.6+621644 & bss & 12 35 38.52 +62 16 43.5  & AGN1 &        & 0.717 & 20.0 & 2 & obs        \\
XBSJ123549.1--395026 & bss & 12 35 49.00 --39 50 24.3  & star &        & -- & 12.4 & 1 & 41         \\
XBSJ123600.7--395217 & bss,hbss & 12 36 00.55 --39 52 15.1  & star &        & -- &  5.8 & 5 & 43         \\
XBSJ123759.6+621102 & bss & 12 37 59.57 +62 11 02.5  & AGN1 &        & 0.910 & 18.4 & 2 & 22,39      \\
XBSJ123800.9+621338 & bss & 12 38 00.92 +62 13 36.0  & AGN1 &        & 0.440 & 18.8 & 2 & 23,39      \\
XBSJ124214.1--112512 & bss & 12 42 13.79 --11 25 10.6  & AGN1 &        & 0.820 & 18.5 & 3 & obs        \\
XBSJ124557.6+022659 & bss & 12 45 57.49 +02 26 57.2  & AGN1 &        & 0.708 & 19.7 & 1 & obs        \\
XBSJ124607.6+022153 & bss & 12 46 07.49 +02 21 53.2  & AGN1 &        & 0.491 & 19.7 & 2 & obs        \\
XBSJ124641.8+022412 & bss,hbss & 12 46 41.70 +02 24 11.3  & AGN1 &        & 0.934 & 17.5 & 2 & 24,39      \\
XBSJ124647.9+020955 & bss & 12 46 47.91 +02 09 54.3  & AGN1 &        & 1.074 & 19.6 & 1 & obs        \\
XBSJ124903.6--061049 & bss & 12 49 03.49 --06 10 47.3  & AGN1 &        & 0.646 & 19.1 & 3 & obs        \\
XBSJ124914.6--060910 & bss & 12 49 14.60 --06 09 09.6  & AGN1 &        & 1.627 & 18.9 & 3 & obs        \\
XBSJ124938.7--060444 & bss & 12 49 38.66 --06 04 44.2  & star &        & -- &  9.7 & 5 & 43         \\
XBSJ124949.4--060722 & bss & 12 49 49.44 --06 07 22.9  & AGN1 &        & 1.053 & 18.6 & 1 & obs        \\
XBSJ125457.2+564940 & bss & 12 54 56.78 +56 49 41.8  & AGN1 &        & 1.261 & 20.3 & 2 & 15,16,39   \\
XBSJ130619.7--233857 & bss & 13 06 19.57 --23 38 56.9  & AGN1 &        & 0.351 & 18.4 & 1 & obs        \\
XBSJ130658.1--234849 & bss & 13 06 58.05 --23 48 47.3  & AGN1 &        & 0.375 & 18.4 & 3 & obs        \\
XBSJ132038.0+341124 & bss,hbss & 13 20 37.88 +34 11 26.2  & AGN1 &        & 0.065 & 16.0 & 2 & 42         \\
XBSJ132052.5+341742 & bss & 13 20 52.56 +34 17 44.1  & AGN1 &        & 0.844 & 21.0 & 2 & 42         \\
XBSJ132101.6+340656 & bss & 13 21 01.43 +34 06 58.0  & AGN1 &        & 0.335 & 18.6 & 2 & 42         \\
\hline
\end{tabular}
\end{table*}
\newpage
\addtocounter{table}{-1}
\begin{table*}
\caption{continue}
\begin{tabular}{l c l l l l r r c}
name & Sample & Optical position & Class & flag class & z & mag & flag mag & reference \\ 
     &        &     (J2000)      &       &      &   &     &  &  \\
\hline
XBSJ132105.5+341459 & bss & 13 21 05.52 +34 15 01.0  & AGN1 &        & 0.452 & 20.3 & 2 & 42         \\
XBSJ133023.8+241707 & bss & 13 30 23.77 +24 17 08.5  & AGN1 &        & 1.438 & 19.3 & 2 & obs        \\
XBSJ133026.6+241520 & bss & 13 30 26.53 +24 15 21.8  & BL &        & 0.460 & 19.2 & 2 & obs        \\
XBSJ133321.2+503102 & bss & 13 33 21.36 +50 31 06.2  & star &        & -- & 11.1 & 5 & obs        \\
XBSJ133626.9--342636 & bss & 13 36 27.00 --34 26 33.0  & star &        & -- & 13.4 & 4 & 41         \\
XBSJ133807.5+242411 & bss & 13 38 07.52 +24 24 11.7  & AGN1 &        & 0.631 & 18.0 & 2 & obs        \\
XBSJ133942.6--315004 & bss,hbss & 13 39 42.47 --31 50 03.0  & AGN1 & 1       & 0.114 & 16.8 & 4 & obs        \\
XBSJ134656.7+580315 & hbss & 13 46 56.75 +58 03 15.7  & AGN2 & 1      3 & 0.373 & 18.3 & 2 & obs        \\
XBSJ134732.0+582103 & bss & 13 47 31.89 +58 21 03.7  & star &        & -- & 14.3 & 2 & obs        \\
XBSJ134749.9+582111 & bss,hbss & 13 47 49.82 +58 21 09.6  & AGN1 &        & 0.646 & 16.7 & 2 & 25,39      \\
XBSJ140100.0--110942 & bss & 14 00 59.93 --11 09 40.8  & AGN1 & 1       & 0.164 & 18.7 & 1 & obs        \\
XBSJ140102.0--111224 & bss,hbss & 14 01 01.83 --11 12 22.8  & AGN1 &       3 & 0.037 & 14.8 & 4 & obs        \\
XBSJ140113.4+024016 & hbss & 14 01 13.32 +02 40 18.8  & AGN1 &        & 0.631 & 21.5 & 1 & obs        \\
XBSJ140127.7+025605 & bss,hbss & 14 01 27.70 +02 56 06.8  & AGN1 &        & 0.265 & 19.3 & 2 & 39         \\
XBSJ140219.6--110458 & bss & 14 02 19.60 --11 04 58.9  & star &        & -- &  8.5 & 5 & 43         \\
XBSJ140936.9+261632 & bss & 14 09 36.88 +26 16 32.3  & star &        & -- & 15.8 & 2 & obs        \\
XBSJ141235.8--030909 & bss & 14 12 35.56 --03 09 09.2  & AGN2 &        & 0.601 & 20.9 & 1 & obs        \\
XBSJ141531.5+113156 & bss,hbss & 14 15 31.48 +11 31 57.3  & AGN1 &        & 0.257 & 18.2 & 2 & 26         \\
XBSJ141722.6+251335 & bss & 14 17 22.53 +25 13 38.2  & AGN1 &    2    & 0.560 & 19.5 & 2 & 27         \\
XBSJ141736.3+523028 & bss & 14 17 35.95 +52 30 30.0  & AGN1 &        & 0.985 & 20.0 & 2 & 28         \\
XBSJ141809.1+250040 & bss & 14 18 08.91 +25 00 42.0  & AGN1 &    2    & 0.727 & 19.4 & 2 & 29         \\
XBSJ141830.5+251052 & bss,hbss & 14 18 30.63 +25 10 53.3  & CL &        & 0.296 & 16.6 & 2 & 30         \\
XBSJ142741.8+423335 & hbss & 14 27 41.62 +42 33 38.1  & AGN2 & 1       & 0.142 & 18.7 & 2 & obs        \\
XBSJ142800.1+424409 & bss & 14 28 00.16 +42 44 11.9  & star &        & -- & 16.5 & 4 & 26         \\
XBSJ142901.2+423048 & bss & 14 29 01.50 +42 30 54.0  & star &        & -- &  9.1 & 5 & 43         \\
XBSJ143835.1+642928 & bss,hbss & 14 38 34.72 +64 29 31.1  & AGN2 & 1      3 & 0.118 & 18.5 & 2 & obs        \\
XBSJ143911.2+640526 & hbss & 14 39 10.72 +64 05 28.9  & AGN2 & 1      3 & 0.113 & 18.2 & 3 & obs        \\
XBSJ143923.1+640912 & bss & 14 39 23.15 +64 09 13.2  & star &        & -- &  7.6 & 5 & 43         \\
XBSJ144937.5+090826 & bss & 14 49 36.61 --09 08 29.6 $^1$ & AGN1 &        & 1.260 & 19.3 & 1 & obs        \\
XBSJ145857.1--313535 & bss & 14 58 57.04 --31 35 37.6  & AGN1 &        & 1.045 & 19.9 & 1 & obs        \\
XBSJ150428.3+101856 & bss & 15 04 28.40 +10 18 56.6  & AGN1 &    2    & 1.000 & 17.7 & 2 & 31         \\
XBSJ151815.0+060851 & bss & 15 18 14.93 +06 08 53.9  & AGN1 &        & 1.294 & 20.0 & 1 & obs        \\
XBSJ151832.3+062357 & bss & 15 18 32.22 +06 23 58.8  & CL &       3 & 0.104 & 16.1 & 2 & obs        \\
XBSJ153156.6--082610 & bss & 15 31 56.60 --08 26 09.1  & star &        & -- &  8.0 & 3 & 40         \\
XBSJ153205.7--082952 & bss & 15 32 05.64 --08 29 50.7  & AGN1 &        & 1.239 & 19.5 & 3 & 40         \\
XBSJ153419.0+011808 & bss & 15 34 19.13 +01 18 04.5  & AGN1 &        & 1.283 & 18.7 & 1 & obs        \\
XBSJ153452.3+013104 & bss,hbss & 15 34 52.53 +01 31 02.9  & AGN1 &        & 1.435 & 18.7 & 3 & 32         \\
XBSJ153456.1+013033 & bss & 15 34 56.32 +01 30 31.1  & AGN1 &        & 0.310 & 17.1 & 3 & obs        \\
XBSJ160645.9+081525 & bss,hbss & 16 06 45.92 +08 15 25.3  & AGN2 &        & 0.618 & 20.1 & 1 & obs        \\
XBSJ160706.6+075709 & bss & 16 07 06.60 +07 57 09.7  & AGN1 &        & 0.233 & 18.7 & 2 & 42,39      \\
XBSJ160731.5+081202 & bss & 16 07 31.61 +08 12 03.4  & AGN1 &        & 0.226 & 19.9 & 2 & 42         \\
XBSJ161820.7+124116 & hbss & 16 18 20.82 +12 41 15.4  & AGN2 & 1       & 0.361 & 19.7 & 2 & obs        \\
XBSJ161825.4+124145 & bss & 16 18 25.56 +12 41 46.7  & AGN1 &        & 0.396 & 19.8 & 2 & obs        \\
XBSJ162813.9+780342 & bss & 16 28 13.40 +78 03 38.2  & AGN1 &    2    & 0.640 & 17.2 & 3 & 33         \\
XBSJ162911.1+780442 & bss & 16 29 10.57 +78 04 39.1  & star &        & -- & 13.0 & 5 & obs        \\
XBSJ162944.8+781128 & bss & 16 29 44.75 +78 11 26.3  & star &        & -- & 16.1 & 4 & obs        \\
XBSJ163141.1+781239 & bss & 16 31 40.84 +78 12 37.4  & AGN1 &        & 0.380 & 18.0 & 3 & 15,16      \\
XBSJ163223.6+052547 & bss & 16 32 23.50 +05 25 44.0  & AGN1 &        & 0.146 & 18.6 & 3 & obs        \\
XBSJ163309.8+571039 & bss & 16 33 09.61 +57 10 41.5  & AGN1 &        & 0.288 & 17.6 & 3 & 15,16      \\
XBSJ163332.3+570520 & bss & 16 33 31.94 +57 05 19.9  & AGN1 & 1       & 0.386 & 18.5 & 3 & 15         \\
XBSJ163427.5+781002 & bss & 16 34 27.40 +78 10 02.7  & AGN1 &        & 0.376 & 19.4 & 3 & 15,16      \\
XBSJ164237.9+030014 & bss & 16 42 37.78 +03 00 11.5  & AGN1 &        & 1.338 & 18.0 & 1 & obs        \\
XBSJ165313.3+021645 & bss & 16 53 13.30 +02 16 46.4  & star &        & -- & 13.6 & 4 & 41         \\
XBSJ165425.3+142159 & bss,hbss & 16 54 25.36 +14 21 59.4  & AGN1 &        & 0.178 & 17.3 & 4 & obs        \\
XBSJ165448.5+141311 & bss,hbss & 16 54 48.62 +14 13 12.2  & AGN1 &       3 & 0.320 & 18.6 & 3 & obs        \\
XBSJ165710.5+352024 & bss & 16 57 10.50 +35 20 24.8  & star &        & -- & 13.7 & 2 & obs        \\
\hline
\end{tabular}
\end{table*}
\newpage
\addtocounter{table}{-1}
\begin{table*}
\caption{continue}
\begin{tabular}{l c l l l l r r c}
name & Sample & Optical position & Class & flag class & z & mag & flag mag & reference \\ 
     &        &     (J2000)      &       &      &   &     &  &  \\
\hline
XBSJ172230.6+341344 & bss & 17 22 30.87 +34 13 40.0  & AGN1 &        & 0.425 & 19.2 & 3 & obs        \\
XBSJ185518.7--462504 & bss & 18 55 18.63 --46 25 04.6  & AGN1 &        & 0.788 & 18.0 & 3 & obs        \\
XBSJ185613.7--462239 & bss & 18 56 13.84 --46 22 37.8  & AGN1 &        & 0.768 & 19.6 & 1 & obs        \\
XBSJ193138.9--725115 & bss & 19 31 39.33 --72 51 15.3  & AGN1 &        & 0.701 & 20.0 & 3 & obs        \\
XBSJ193248.8--723355 & bss,hbss & 19 32 48.56 --72 33 53.0  & AGN2 & 1       & 0.287 & 18.8 & 3 & obs        \\
XBSJ204043.4--004548 & bss,hbss & 20 40 43.48 --00 45 49.6  & AGN2 &        & 0.615 & 21.2 & 1 & obs        \\
XBSJ204159.2--321439 & bss & 20 41 59.20 --32 14 40.3  & AGN1 &        & 0.738 & 19.8 & 1 & obs        \\
XBSJ204204.1--321601 & bss & 20 42 04.16 --32 16 02.1  & AGN1 &        & 0.384 & 20.1 & 3 & obs        \\
XBSJ204208.2--323523 & bss & 20 42 08.14 --32 35 23.2  & AGN1 &        & 1.184 & 20.9 & 1 & obs        \\
XBSJ204548.4--025234 & bss & 20 45 48.41 --02 52 34.7  & AGN1 &        & 2.188 & 18.1 & 3 & obs        \\
XBSJ205411.9--160804 & bss & 20 54 12.04 --16 08 03.0  & AGN1 &        & 1.466 & 17.7 & 3 & obs        \\
XBSJ205429.9--154937 & bss & 20 54 30.10 --15 49 35.8  & AGN1 &        & 1.297 & 18.6 & 3 & obs        \\
XBSJ205635.7--044717 & bss,hbss & 20 56 35.63 --04 47 17.1  & AGN1 &        & 0.217 & 17.3 & 3 & obs        \\
XBSJ205829.9--423634 & bss,hbss & 20 58 29.89 --42 36 34.3  & AGN1 &        & 0.232 & 18.3 & 3 & obs        \\
XBSJ205847.0--423704 & bss & 20 58 47.01 --42 37 04.6  & star &        & -- & 14.2 & 4 & 41         \\
XBSJ210325.4--112011 & bss & 21 03 25.31 --11 20 11.2  & AGN1 &        & 0.720 & 20.1 & 3 & obs        \\
XBSJ210355.3--121858 & bss & 21 03 55.20 --12 18 58.4  & AGN1 &        & 0.792 & 19.5 & 3 & obs        \\
XBSJ212635.8--445046 & bss & 21 26 35.84 --44 50 47.7  & star &        & -- &  7.9 & 5 & 43         \\
XBSJ212759.5--443924 & bss & 21 27 59.79 --44 39 24.6  & AGN1 &        & 0.860 & 21.1 & 1 & obs        \\
XBSJ213002.3--153414 & bss,hbss & 21 30 02.31 --15 34 12.9  & AGN1 &        & 0.562 & 17.3 & 3 & obs        \\
XBSJ213719.6--433347 & bss & 21 37 19.86 --43 33 47.9  & AGN1 &        & 0.793 & 20.8 & 3 & obs        \\
XBSJ213729.7--423601 & bss & 21 37 29.87 --42 36 00.3  & AGN1 &        & 0.664 & 19.9 & 1 & obs        \\
XBSJ213733.2--434800 & bss & 21 37 33.52 --43 48 00.8  & AGN1 &        & 0.427 & 20.0 & 3 & obs        \\
XBSJ213757.6--422334 & bss & 21 37 58.20 --42 23 30.1  & AGN1 &        & 0.364 & 18.8 & 1 & obs        \\
XBSJ213820.2--142536 & bss,hbss & 21 38 20.19 --14 25 32.8  & AGN1 &        & 0.369 & 19.0 & 3 & obs        \\
XBSJ213824.0--423019 & bss & 21 38 23.98 --42 30 16.1  & AGN1 &        & 0.257 & 17.5 & 4 & 34         \\
XBSJ213829.8--423958 & bss & 21 38 29.89 --42 39 57.5  & AGN1 &        & 1.469 & 17.7 & 3 & 35         \\
XBSJ213840.5--424241 & bss & 21 38 40.54 --42 42 40.1  & star &        & -- &  9.3 & 5 & 43         \\
XBSJ213852.2--434714 & bss & 21 38 52.52 --43 47 15.3  & AGN1 &        & 0.461 & 18.5 & 3 & 36         \\
XBSJ214041.4--234720 & bss,hbss & 21 40 41.46 --23 47 19.1  & AGN1 &        & 0.490 & 18.4 & 3 & obs        \\
XBSJ215244.2--302407 & bss & 21 52 44.23 --30 24 05.7  & AGN1 &        & 0.539 & 17.9 & 3 & obs        \\
XBSJ215323.7+173018 & bss & 21 53 23.67 +17 30 20.6  & star &        & -- & 14.5 & 4 & obs        \\
XBSJ220320.8+184930 & bss & 22 03 21.02 +18 49 31.6  & AGN1 &       3 & 0.309 & 20.1 & 1 & obs        \\
XBSJ220446.8--014535 & bss & 22 04 46.89 --01 45 34.7  & AGN1 &        & 0.540 & 21.5 & 2 & obs        \\
XBSJ220601.5--015346 & bss,hbss & 22 06 01.45 --01 53 45.1  & AGN1 &        & 0.211 & 20.1 & 3 & obs        \\
XBSJ221623.3--174317 & bss & 22 16 23.50 --17 43 16.1  & AGN1 &        & 0.754 & 19.6 & 2 & 40         \\
XBSJ221722.4--082018 & bss & 22 17 22.39 --08 20 17.0  & AGN1 &        & 1.160 & 19.9 & 1 & obs        \\
XBSJ221729.3--081154 & bss & 22 17 29.40 --08 11 55.0  & AGN1 &        & 1.008 & 19.7 & 3 & obs        \\
XBSJ221750.4--083210 & bss & 22 17 50.35 --08 32 10.2  & star &        & -- & 15.6 & 4 & obs        \\
XBSJ221821.9--081332 & bss & 22 18 21.87 --08 13 29.8  & AGN1 &        & 0.803 & 19.2 & 3 & obs        \\
XBSJ221951.6+120123 & bss & 22 19 51.52 +12 01 20.9  & AGN2 &        & 0.532 & 20.0 & 2 & obs        \\
XBSJ222852.2--050915 & bss & 22 28 52.22 --05 09 13.3  & star &        & -- &  9.6 & 5 & 40         \\
XBSJ223547.9--255836 & bss & 22 35 48.14 --25 58 35.2  & AGN1 &        & 0.304 & 19.1 & 3 & obs        \\
XBSJ223555.0--255833 & bss & 22 35 55.09 --25 58 33.0  & AGN1 &        & 1.800 & 18.5 & 3 & obs        \\
XBSJ223949.8+080926 & bss & 22 39 50.21 +08 09 29.0  & AGN1 &        & 1.406 & 19.1 & 3 & obs        \\
XBSJ224756.6--642721 & bss & 22 47 56.61 --64 27 18.5  & AGN1 &        & 0.598 & 18.5 & 1 & obs        \\
XBSJ224833.3--511900 & bss & 22 48 33.30 --51 19 00.9  & star &        & -- &  3.5 & 5 & 43         \\
XBSJ224846.6--505929 & bss & 22 48 46.58 --50 59 28.1  & star &        & -- & 13.0 & 6 & 43         \\
XBSJ225025.1--643225 & bss & 22 50 25.32 --64 32 26.2  & AGN1 &        & 1.206 & 19.7 & 3 & obs        \\
XBSJ225050.2--642900 & bss & 22 50 50.51 --64 29 03.0  & AGN1 &        & 1.251 & 18.5 & 3 & obs        \\
XBSJ225118.0--175951 & bss & 22 51 18.02 --17 59 48.9  & AGN1 &        & 0.172 & 19.0 & 3 & obs        \\
XBSJ225349.6--172137 & bss & 22 53 49.64 --17 21 36.4  & star &        & -- & 16.1 & 4 & obs        \\
XBSJ230400.4--083755 & bss & 23 04 00.59 --08 37 53.8  & AGN1 &        & 0.411 & 19.5 & 3 & obs        \\
XBSJ230401.0+031519 & bss & 23 04 01.18 +03 15 18.5  & AGN1 & 1      3 & 0.036 & 14.0 & 4 & obs        \\
XBSJ230408.2+031820 & bss & 23 04 08.40 +03 18 20.9  & star &        & -- & 11.5 & 5 & 43         \\
XBSJ230434.1+122728 & bss & 23 04 34.25 +12 27 26.2  & AGN1 & 1      3 & 0.232 & 18.3 & 3 & obs        \\
\hline
\end{tabular}
\end{table*}
\newpage
\addtocounter{table}{-1}
\begin{table*}
\caption{continue}
\begin{tabular}{l c l l l l r r c}
name & Sample & Optical position & Class & flag class & z & mag & flag mag & reference \\ 
     &        &     (J2000)      &       &      &   &     &  &  \\
\hline
XBSJ230443.8+121636 & bss & 23 04 43.75 +12 16 36.6  & AGN1 &        & 1.405 & 19.8 & 3 & obs        \\
XBSJ230459.6+121205 & bss & 23 04 59.64 +12 12 05.8  & AGN1 &       3 & 0.560 & 20.7 & 1 & obs        \\
XBSJ230522.1+122121 & bss & 23 05 22.14 +12 21 20.2  & AGN2 &        & 0.326 & 19.5 & 3 & obs        \\
XBSJ231342.5--423210 & bss & 23 13 42.53 --42 32 09.2  & AGN1 &        & 0.973 & 19.1 & 3 & obs        \\
XBSJ231541.2--424125 & bss & 23 15 41.37 --42 41 26.4  & star &        & -- &  9.9 & 5 & 43         \\
XBSJ231546.5--590313 & bss & 23 15 46.76 --59 03 14.5  & AGN2 & 1       & 0.045 & 14.0 & 1 & 37         \\
XBSJ231553.0--423800 & bss & 23 15 52.97 --42 38 00.0  & star &        & -- & 13.9 & 4 & 41         \\
XBSJ231601.7--424038 & bss & 23 16 01.66 --42 40 38.1  & AGN1 &        & 0.383 & 19.2 & 1 & obs        \\
XBSJ233325.7--152240 & bss & 23 33 26.05 --15 22 37.7  & star &        & -- & 13.9 & 4 & obs        \\
XBSJ233421.9--151219 & bss & 23 34 22.14 --15 12 16.9  & AGN1 &        & 0.992 & 19.5 & 3 & obs        \\
XBSJ235032.3+363156 & bss & 23 50 32.35 +36 32 00.2  & star &        & -- & 13.1 & 1 & obs        \\
XBSJ235036.9+362204 & bss & 23 50 36.97 +36 22 05.7  & BL &        & 0.317 & 17.6 & 3 & 38         \\
\hline
\end{tabular}

\caption{The complete list of XBS sources with a spectroscopic classification.
{\bf column~1}. Name;
{\bf column~2}. The sample to which the source belongs (BSS or HBSS);
{\bf column~3}. The position of the optical counterpart ($^1$ = in this object the offset
between the optical and the X-ray position, given in Della Ceca et al. 2004, is 15\arcsec i.e. significantly
larger than the X-ray positional error. In this particular case the X-ray position 
was wrongly determined. To find the correct optical counterpart we have used the improved 
X-ray position found in the 2XMM catalogue); 
{\bf column~4}. The spectral classification (see text for details);
{\bf column~5}. A classification flag (1 = classification based on the X-ray 
analysis; 2 = no spectrum or table with relevant lines property found in the 
literature but only a classification; 3 = classification different from that
presented in Della Ceca et al. 2004);
{\bf column~6}. The redshift; 
{\bf column~7}. The magnitude (mostly in a red filter);
{\bf column~8}. A flag indicating the magnitude filter (1 = R magnitude; 2 = 
r magnitude; 3 = APM red magnitude; 4 = APM red magnitude corrected according 
to the relation discussed in the text; 5 = V magnitude; 6 = B magnitude);
{\bf column~9}. The origin of the spectral data 
used to classify the source (
     obs = our own observations ;
      1 =   Fiore et al. 2003     ;   
      2 =   Bechtold et al. 2002  ;   
      3 =   Schneider et al. 2003 ;   
      4 =   Cristiani et al. 1995 ;   
      5 =   Burbidge 1999        ;   
      6 =   La Franca et al. 1992 ;   
      7 =   Fiore et al. 2000    ;   
      8 =   Zitelli et al. 1992   ;   
      9 =   Croom et al. 2001     ;   
     10 =   Mignoli et al. 2005   ;   
     11 =   Meyer et al. 2001          ;   
     12 =   Morris et al. 1991    ;   
     13 =   Wei et al. 1999       ;   
     14 =   Nagao et al. 2001     ;   
     15 =   Mason et al. 2000     ;   
     16 =   Puchnarewicz etal 1997  ;   
     17 =   Lehmann et al. 2000   ;   
     18 =   Ho et al. 1997        ;   
     19 =   Ho et al. 1995        ;   
     20 =   Ebeling et al. 2001   ;   
     21 =   Cagnoni et al. 2001   ;   
     22 =   Vanden Berk et al. 2000 ;   
     23 =   Liu et al. 1999       ;   
     24 =   Hewett et al. 1991    ;   
     25 =   Bade et al. 1995      ;   
     26 =   Boyle et al. 1997     ;   
     27 =   Stocke et al. 1983    ;   
     28 =   Hammer et al. 1995    ;   
     29 =   Burbidge et al. 2002  ;   
     30 =   Romer et al. 2000     ;   
     31 =   Arnaud et al. 1985    ;   
     32 =   Baldwin et al. 1989   ;   
     33 =   Pietsch \& Arp 2001   ;   
     34 =   Hewit et al. 1995     ;   
     35 =   Morris et al. 1991b   ;   
     36 =   Cristiani et al. 1990 ;   
     37 =   Kewley et al. 2001    ;   
     38 =   Perlman et al. 1998   ;   
     39 =   SDSS rel.5 (http://cas.sdss.org/dr5/);
     40 =   Barcons et al. 2007  ;
     41 =   L\'opez-Santiago et al. 2007;
     42 =   Unpublished spectra taken 
     through the AXIS collaboration (http://venus.ifca.unican.es/~xray/AXIS/) ;
     43 =   classification from Simbad
)
}
\end{table*}

\appendix

\end{document}